\documentclass[twocol]{ametsocV6.1}

\usepackage{graphicx} 

\usepackage{tabularx}
\usepackage{booktabs, amsmath}
\usepackage{amssymb}
\usepackage{siunitx}
\usepackage{booktabs}
\usepackage{appendix}
\usepackage{pdfpages}

\title{FastNet: Improving the physical consistency of machine-learning weather prediction models through loss function design}

\date{September 2025}

\authors{Tom Dunstan*,\aff{a}\correspondingauthor{Tom Dunstan, tom.dunstan@metoffice.gov.uk}
Oliver Strickson*,\aff{b}\correspondingauthor{Oliver Strickson, ostrickson@turing.ac.uk}\thanks{* Equal contributions} 
Thusal Bennett*,\aff{a} 
Jack Bowyer*,\aff{a} 
Matthew Burnand*,\aff{a} 
James Chappell*,\aff{a} 
Alejandro Coca-Castro*,\aff{b} 
Kirstine Ida Dale*, \aff{a}
Eric G. Daub*,\aff{b} 
Noushin Eftekhari*,\aff{b} 
Manvendra Janmaijaya*,\aff{b} 
Jon Lillis*,\aff{a} 
David Salvador-Jasin*,\aff{b} 
Nathan Simpson*,\aff{b}
Ryan Sze-Yin Chan*,\aff{b} 
Mohamad Elmasri*, \aff{c} 
Lydia Allegranza France*,\aff{b} 
Sam Madge*,\aff{a} 
Levan Bokeria,\aff{b}
Hannah Brown,\aff{g} 
Tom Dodds,\aff{h} 
Anna-Louise Ellis,\aff{a} 
David Llewellyn-Jones,\aff{b}
Theo McCaie,\aff{a} 
Sophia Moreton,\aff{b} 
Tom Potter,\aff{a} 
James Robinson,\aff{b} 
Adam A. Scaife,\aff{d,e} 
Iain Stenson,\aff{b} 
David Walters,\aff{a} 
Karina Bett-Williams,\aff{a,f}
Louisa van Zeeland,\aff{b} 
Peter Yatsyshin,\aff{b} 
and J. Scott Hosking\aff{b,i}
}

\affiliation{\aff{a}{Met Office, Exeter, UK}\\
\aff{b}{The Alan Turing Institute, London, UK}\\
\aff{c}{Carleton University, Ottawa, Canada}\\
\aff{d}{Met Office Hadley Centre, Exeter, UK}\\
\aff{e}{Department of Mathematics and Statistics, University of Exeter, Exeter, UK}\\
\aff{f}{Global Systems Institute, University of Exeter, Exeter, UK}\\
\aff{g}{UK Hydrographic Office}\\
\aff{h}{Office for National Statistics, UK}\\
\aff{i}{British Antarctic Survey, Cambridge, UK}
}

\abstract{Machine learning weather prediction (MLWP) models have demonstrated remarkable potential in delivering accurate forecasts at significantly reduced computational cost compared to traditional numerical weather prediction (NWP) systems. However, challenges remain in ensuring the physical consistency of MLWP outputs, particularly in deterministic settings. This study presents FastNet, a graph neural network (GNN)-based global prediction model, and investigates the impact of alternative loss function designs on improving the physical realism of its forecasts.
We explore three key modifications to the standard mean squared error (MSE) loss: (1) a modified spherical harmonic (MSH) loss that penalises spectral amplitude errors to reduce blurring and enhance small-scale structure retention; (2) inclusion of horizontal gradient terms in the loss to suppress non-physical artefacts; and (3) an alternative wind representation that decouples speed and direction to better capture extreme wind events.
Results show that while the MSH and gradient-based losses \textit{alone} may slightly degrade RMSE scores, when trained in combination the model exhibits very similar MSE performance to an MSE-trained model while at the same time significantly improving spectral fidelity and physical consistency. The alternative wind representation further improves wind speed accuracy and reduces directional bias. Collectively, these findings highlight the importance of loss function design as a mechanism for embedding domain knowledge into MLWP models and advancing their operational readiness.}

\begin{document}
\includepdf{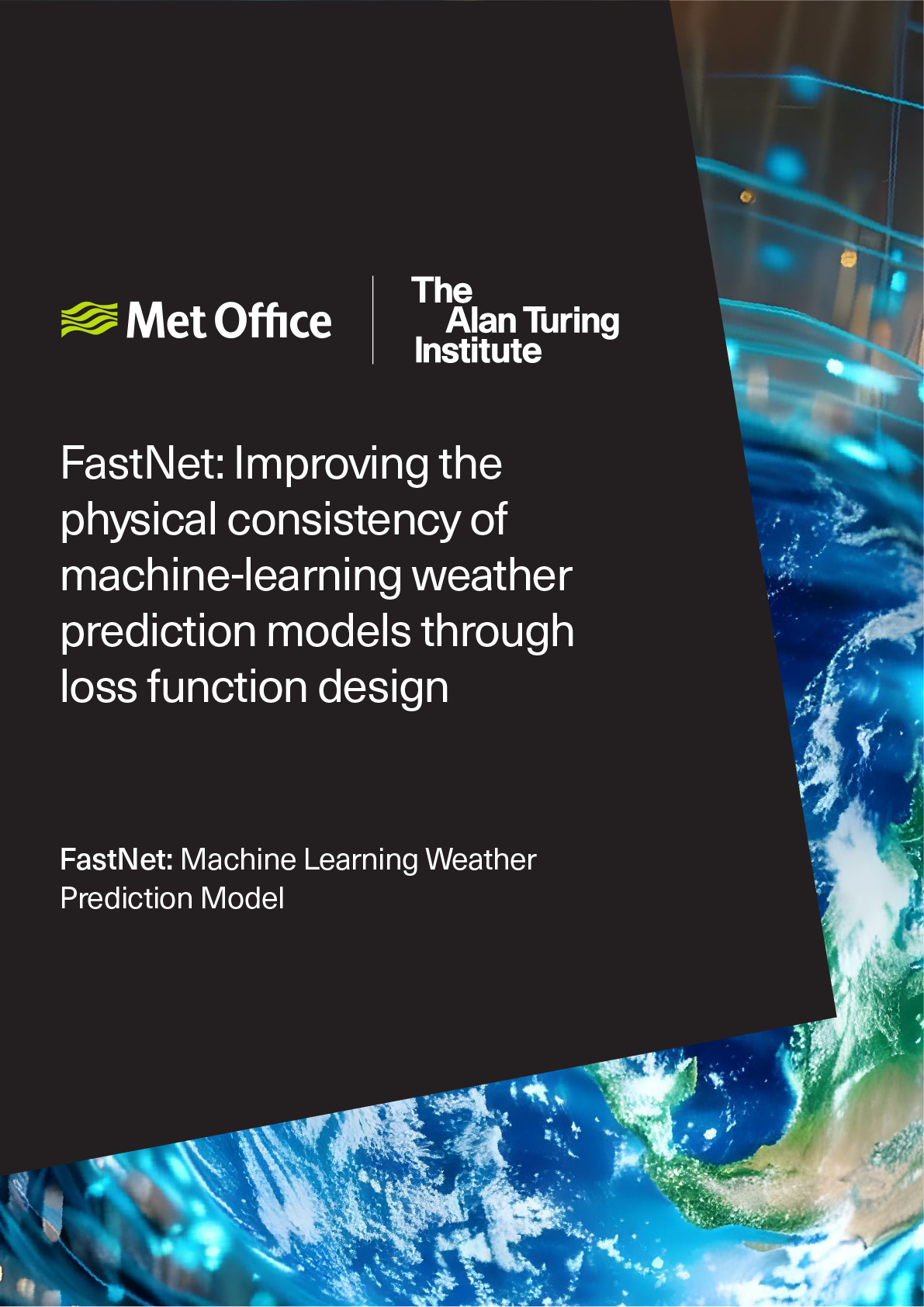}

\maketitle

\section{Introduction}
Within a time frame of only a few years, Machine Learning (ML) based weather prediction has risen from speculative, small-scale research projects to established components of many national met services' (NMS) plans for future operational products. The potential benefits of Machine Learning Weather Prediction (MLWP) models are clear: bringing a step change in accuracy on many standard error metrics alongside an orders-of-magnitude reduction in the time and cost needed to produce a forecast. Beyond this tremendous potential, however, there remains a substantial gap between the outputs currently available from MLWP models (temporal and spatial resolution and the range of processes represented) and the outputs available from current operational Numerical Weather Prediction (NWP) models. Whether or not this gap can be closed with purely data-driven models or whether a blend of traditional and data-driven modelling will continue to be necessary is an active area of research. Regardless of the answer, a key factor in closing this gap will come from improving the physical consistency of MLWP output and moving beyond standard error metrics such as Root Mean Squared Error (RMSE). Achieving this will broaden the range of downstream uses for MLWP models, increase confidence and trust in their output, and improve their ability to generalise to unseen conditions.

We focus here on deterministic MLWP models. Whilst ensembles are of growing importance even at short lead times, good deterministic model characteristics are important in forming a good ensemble. We also make the assumption here that initial conditions are available from an existing analysis, and so the important challenge of data assimilation, or of working directly from observations, is set to one side for the purposes of this work.

Early attempts at pure ML-based weather models adapted image processing architectures such as Convolutional Neural Nets (CNNs) \citep{rasp_weatherbench_2020,rasp_datadriven_2021}. The models were skillful but resolution was relatively coarse and errors were not competitive with the state of the art NWP models at the time. The seminal paper by Keisler \citep{keisler_forecasting_2022}, building on work from \citet{pfaff2021learning}, was the first to demonstrate the true potential of the data-driven approach; producing a model that was competitive with leading NWP systems on key error metrics and with training and inference costs that were orders of magnitude lower than those required by NWP. Many of the design choices established in \citet{keisler_forecasting_2022}, such as the encode-process-decode GNN structure, auto-regressive rollout, increment learning, and multi-resolution training still form the basis for many leading MLWP models today.

FourCastNet \citep{pathak_fourcastnet_2022}, using a vision transformer backbone operating in Fourier space, was the first to achieve \ang{0.25} resolution using ERA5 on a lat/lon grid. Pangu Weather \citep{bi_accurate_2023} extended the vision transformer concept to accommodate 3-dimensional atmospheric blocks, and was the first to achieve competitive RMSE scores compared to state of the art NWP models.

In parallel to the transformer-based approach, Graph-based models were further advanced in GraphCast \citep{lam_learning_2023}, building on the work of \citet{keisler_forecasting_2022} to produce a state of the art model that exceeded the performance of the Integrated Forecast System (IFS) from the European Centre for Medium-Range Weather Forecasting (ECMWF) on the majority of standard error metrics. The AIFS model developed by ECMWF \citep{lang_aifs_2024} also followed the encode-process-decode GNN architecture and was the first to exceed IFS performance for all variables.

Alternative approaches such as Spherical Fourier Neural Operators \citep{bonev_spherical_2023, bonev_fourcastnet_2025} and Hybrid General Circulation-NN models \citep{kochkov_neural_2024} have also proven effective for producing competitive prediction models. 


The diversity of approaches described above, all capable of producing competitive forecasts, points to the fact that, beyond certain minimum requirements, the choice of model architecture is not the only, or even the most important, factor affecting performance, but rather
\begin{itemize}
    \item the size and quality of training data (length, consistency, resolution, accuracy)
    \item the formulation of the learning task (including the design of the loss function)
    \item the computational efficiency and scaling characteristics of the model
\end{itemize}
all work in combination to determine a model's effectiveness.

The FastNet model therefore takes as its starting point a GNN-based architecture similar to those described in \citet{pfaff2021learning, keisler_forecasting_2022, lam_learning_2023}, and uses ERA5 reanalysis data for training on a global domain. Our main contribution focuses instead on improving the physical consistency and generalisability of the model through modifications to the loss function.

In particular, we focus on three key areas where deterministic ML models have been found to exhibit biases:
\begin{itemize}
    \item Spectral biases, i.e. loss of power at the small to medium scales (blurring) that compounds with lead time, e.g. \citet{lam_learning_2023}.
    \item A slow bias in the wind speeds, particularly at the extremes \citep{charlton-perez_ai_2024}.
    \item Artefacts that appear first in the horizontal derivatives and then forecast variables at longer lead times \citep{charlton-perez_ai_2024}.
\end{itemize}

Further details of each of these are given in the relevant sections below. The underlying hypothesis for the work presented here is that if the loss landscape following pre-training is relatively flat, or exhibits mode connectivity \citep{wilson_deep_2025}, then multiple solutions may exist that will maintain good performance on the pre-training loss target (grid-point MSE in this case), but where other aspects of the model performance can be introduced or enhanced. Alternatively, it may be that, as with the well understood case of RMSE vs spectral bias (a manifestation of the double penalty effect) there is an unavoidable trade-off between different aspects of model performance, and that choices must be made about the relative importance of each of these aspects for individual use cases.

We provide a brief description of the model architecture, data, and training in Section \ref{sec:fastnet_model}. Further details can be found in a companion paper \citep{Daub_fastnet_paper1}. Baseline model results using the standard MSE loss function are used here only as a reference to present the effects of individual loss function modifications. Alternative loss function experiments are described in Section \ref{sec:alt_loss_funcs} and include: an implementation of the Modified Spherical Harmonic (MSH) loss first described in \citet{subich_fixing_2025}, inclusion of horizontal gradients in the loss function, and an alternative representation of winds. Finally, results from FastNet v1.1: the consolidated model where all loss function modifications are combined, are presented in Section \ref{sec:consolidated_results}. Conclusions are given in Section \ref{sec:conclusions}.

\section{FastNet Model}
\label{sec:fastnet_model}

The FastNet architecture is similar to that described in \citet{lam_learning_2023}, employing an encode $>$ process $>$ decode design that takes a gridded atmospheric state at time $t_0$, encodes this to a latent space on a lower-dimensional mesh, and outputs an increment to the atmospheric state over time interval, $\Delta t$, which is then added to the state at $t_0$ to produce a forecast at $t_1 = t_0 + \Delta t$. This process is applied recursively to produce forecasts at lead times $n \Delta t$, where $n$ is a positive integer.

\subsection{Mesh}
FastNet represents the Earth using a multi-scale icosahedral mesh \citep{dahl2014surface, lam_learning_2023} that balances uniform coverage and computational efficiency. Starting from a base icosahedron with 12 nodes, the mesh is refined by subdividing each edge into equal segments N times until the target angular spacing is reached. For N=6 this yields 40,962 nodes. Edges from all refinement levels are retained in the final graph, so the resulting structure contains fine-scale edges connecting nearby nodes and long-range edges spanning coarser levels. This multi-scale connectivity allows information to propagate globally in only a few message-passing steps, while still capturing local structure without excessive hops. Because the mesh nodes lie on a uniform spherical sampling, FastNet avoids the polar clustering and singularities of regular latitude–longitude grids. We train FastNet at a grid resolution of approxiamtely \ang{1}. For the \ang{1} data, N=5 which produces a mesh with approximately \ang{2} spacing.

\subsection{Encoder}
FastNet’s encoder is a bipartite graph that aggregates information from the input grid onto the icosahedral mesh. Grid to mesh connectivity is based on a $k$-nearest-neighbour (KNN) method, where each grid point is connected to its $k$ closest mesh nodes. We also tested a radius based method that creates connections between grid and mesh nodes based on radial distance from each mesh node, but found only a small effect on performance. More details of this are given in \citet{Daub_fastnet_paper1}.

\subsection{Processor}
FastNet’s core is a 16-layer Interaction Network (a graph neural network) \cite{battaglia2016interaction} that advances latent features on the icosahedral mesh. In each layer, every edge $(i, j)$ computes a message from node $i$ to node $j$ using a multilayer perceptron (MLP) that takes the latent state of node $i$ concatenated with edge attributes (e.g., length and orientation). Messages from all neighbours across all refinement levels are aggregated at each node, and a node-wise MLP updates the node’s latent state. A residual skip-connection is applied to the edge features at every iteration. Retaining edges from all levels allows local information to flow along fine edges and global signals to propagate efficiently along coarse edges. After 16 layers, the node encodes multi-scale information across the globe.

\subsection{Decoder}
The decoder maps latent features on the mesh back to the target grid. Specifically, each grid node identifies its three nearest mesh nodes using Haversine distance. Directed edges are drawn from these nodes to the grid node, and each mesh node passes its latent state through an MLP to produce a message. The grid node then concatenates messages from its three input nodes and applies another MLP to produce a vector of atmospheric variables at that grid point. Because FastNet adopts a residual formulation, the decoder’s output is an increment added to the current atmospheric state, so the model learns only the six-hour change rather than the full field.
By training the encoder, processor, and decoder end-to-end, FastNet jointly optimizes feature extraction, temporal evolution, and spatial interpolation under a single objective that minimizes the loss on all physical variables. This unified design enables the model to extract multi-scale information efficiently and produce skilful forecasts across all targeted lead times.

\subsection{Data}   
    FastNet uses the Copernicus ERA5 reanalysis dataset produced by ECMWF \citep{hersbach_era5_2020}. It takes initial conditions and produces forecasts on ERA5 reduced Gaussian grids. In this study, we use ERA5 data at O96 (an octohedral reduced Gaussian grid), which correspond approximately to 104 km, or \ang{1}, grid spacing. The use of reduced Gaussian grids provides a considerable reduction to the total number of grid points (and computation) compared to regular latitude-longitude grids of similar resolutions. For instance, the N320 reduced Gaussian grid has 542,080 grid points, while the equivalent \ang{0.25} latitude-longitude grid has 1,038,240 grid points. 

Inputs of the trained models include 13-pressure levels of standard upper air variables, surface variables and forcing surface variables such as land-sea mask, orography and temporal ancillary fields. The full list of input and output fields of FastNet is shown in Table~\ref{tab:fastnet_variables}.  

The training data set comprises analyses taken at 6hr intervals (00Z, 06Z, 12Z, and 18Z) for years spanning 1980-2020, with validation on data from 2021. Results presented here use data from 2022.

\begin{table*}[t!]
\centering
\caption{Variables used in FastNet training. }
\renewcommand{\arraystretch}{1.5} 
\setlength{\extrarowheight}{2pt} 
\begin{tabularx}{\textwidth}{|X|X|X|}
\hline
\textbf{Variable} & \textbf{Level type} & \textbf{Input/Output} \\
\hline
Geopotential \newline Horizontal wind components \newline Specific humidity \newline Temperature &Pressure level: 50, 100, 150, 200, 250, 300, 400, 500, 600, 700, 850, 925, 1000 &Both \\
\hline
Surface pressure \newline Mean sea-level pressure \newline Skin temperature \newline 2m temperature \newline 2m dewpoint temperature \newline  10m horizontal wind components \newline &Surface &Both \\
\hline
Land-sea mask \newline Orography \newline Standard deviation of sub-grid orography \newline Slope of sub-scale orography \newline Top-of-atmosphere solar radiation \newline Cos/sin of time of day/day of year \newline Cos/sin of lat/lon &Surface &Input \\
\hline
\end{tabularx}
\label{tab:fastnet_variables}
\end{table*}

\subsection{Training}

The training objective of the baseline FastNet model is a mean squared-error (MSE) between the target output $x$ and forecast output $\hat{x}$ that combines errors by forecast variable and lead time.

\begin{equation}
\mathcal{L}_{MSE}=\frac{1}{\left|D_{\mathrm{batch}}\right|}\sum_{d_0\in D_{\mathrm{batch}}}{\frac{1}{T_{\mathrm{train}}}\sum_{\tau\in1:T_{\mathrm{train}}}\sum_{j\in J}{s_jw_j}}\left({\hat{x}}_j^{d_0+\tau}\ -\ x_j^{d_0+\tau}\ \right)^{2\ }
\end{equation}

\noindent where, 

\begin{itemize}
    \item $\tau \in 1 : T_\mathrm{train}$ are the lead times being forecast by the model.
    \item $d_0 \in D_\mathrm{batch}$ represent initialisation states within a given batch.
    \item $j \in J$ represents the variable index.
    \item $s_j$ represent the per-variable-level inverse variance of time differences.
    \item $w_j$ represent the per-variable-level pressure or surface weight.
    \item ${\hat{x}}_j^{d_0+\tau}$ and $x_j^{d_0+\tau}$ are the forecast and target values for some variable-level and lead time.
\end{itemize}

\noindent Training FastNet is performed in two phases: pre-training on 6hr increments only with a maximal learning rate schedule, followed by fine-tuning at a low learning rate where the number of roll-out steps is gradually increased up to a maximum of 12 steps (3 days). The pre-training phase involved 150 epochs with cosine annealing and a peak learning rate of $\SI{1e-3}{}
$.

\section{Alternative loss functions}
\label{sec:alt_loss_funcs}

\subsection{Spherical harmonic loss}

\noindent A formulation of the modified spherical harmonic (MSH) loss function first described in \cite{subich_fixing_2025} can be optionally used when training the FastNet model. This decomposes each of the scalar fields, $\hat{x}(i, j)$, into complex spherical harmonics as:

\begin{equation}
\hat{x} (i,j) = \sum_k \sum_{l=-k}^{k} \alpha_{\hat{x}} (k,l) \mathbb{Y}^{l}_{k}(i,j),    
\end{equation}

\noindent where $\mathbb{Y}^{l}_{k}$ is the complex-valued spherical harmonic mode defined by total wavenumber $k$ and zonal wavenumber $l$, and $\alpha_x(k,l)$ is the corresponding spectral coefficient. The same decomposition can be performed for both the predictions provided by the model ($\hat{x}$) and the corresponding ground truth ($x$), and an \textit{adjusted} mean squared error (AMSE) per-variable loss term can be defined:

\begin{align}
&\mathrm{AMSE}(\hat{x},x) 
= \sum_k \bigg[
\left( \sqrt{\mathrm{PSD}_k(\hat{x})} - \sqrt{\mathrm{PSD}_k(x)} \right)^2 \notag \\
&+ 2 \cdot \max\left(\mathrm{PSD}_k(\hat{x}), \mathrm{PSD}_k(x)\right)
    \cdot \left(1 - \mathrm{Coh}_k(\hat{x}, x)\right)
\bigg], \label{eq:amse}
\end{align}

\noindent where the power spectral density of a field is calculated via $\mathrm{PSD}_k(x) = \sum_l |\alpha_x(k,l)|^2$, and the coherence of the forecast and ground truth field is 

\begin{equation}
    \mathrm{Coh}_k(\hat{x},x) = \frac{\sum_l \mathbb{Re}  \left[\alpha_{\hat{x}}(k,l) \cdot \alpha_x^*(k,l)\right]}{\sqrt{\mathrm{PSD}_k(\hat{x}) \mathrm{PSD}_k(x)}},
\end{equation}

\noindent where $^*$ represents the complex conjugate. Use of the MSH loss function within training penalises errors in spectral amplitude and therefore encourages the model to produce predictions with improved physical consistency. It additionally helps to minimise the smoothing of fine scale structures often observed in data-driven weather prediction models, particularly at long lead times. It should be noted however that this can have a detrimental effect on RMSE scores as the model is no longer rewarded for ``blurring'' predictions.

\subsubsection{Practical implementation}

\noindent The spherical harmonic transform (SHT) is performed using the \texttt{torch-harmonics} package~\citep{bonev_spherical_2023} which, in its current implementation, requires that the fields to be transformed are defined on a regular latitude-longitude (lat-lon) grid. However, the FastNet model operates on reduced gaussian grids and hence the forecast and target fields must be regridded at each training step. To minimise computational cost, a sparse matrix containing the weights that represent the regridding operation from the reduced gaussian grid to its ``equivalent'' regular lat-lon grid (i.e. $\mathrm{O}96 \rightarrow 1.0^{\circ},\, \mathrm{N}320 \rightarrow 0.25^{\circ}$) is pre-calculated, reducing the regridding operation to a single sparse matrix multiplication for a given batch. 


\subsubsection{Weighting schemes}

\subsubsection*{Per-variable-level weights}

\noindent As for the mean squared error (MSE) formulation, a series of weightings~\citep{lam_learning_2023} are applied to each forecast field when calculating the loss within a single training step. These include a per-variable-level weighting proportional to the pressure level, 

\begin{equation}
    w_j = \frac{P_j}{\sum_j P_j},
\end{equation}

\noindent for variables defined on multiple pressure levels, and use fixed weights for surface variables:

\begin{equation}
w_j = 
\begin{cases}
1       & \text{if } \texttt{T2M}, \\
0.1     & \text{otherwise}.
\end{cases}
\end{equation}

\noindent We also include a weight corresponding to the per-variable-level inverse variance of time differences, 

\begin{equation}
    s_j = \mathbb{V}_{i,t} \left[x^{t+1}_{i,j} - x^t_{i,j} \right]^{-1},
\end{equation}

\noindent such that each variable --- indexed by $j$ --- has a corresponding per-variable-level weight, $w_j \times s_j$.

In FastNet, the spherical harmonic loss applies an \textit{additional} per-variable weight to balance contributions across variables before the main per-variable-level weights. This correction accounts for disparate PSD (and thus AMSE) scales: for instance, geopotential spectra span more than six orders of magnitude on the $\mathrm{O}96$ grid, whereas specific humidity spans only three. Without this adjustment, the MSH loss tends to overweight $v$-winds and humidity while underweighting temperature and geopotential. We therefore scale the AMSE contribution from each variable-level combination by the inverse of its ground-truth AMSE, calculated from comparing consecutive 6-hour time-steps over a given year:

\begin{equation}
    \beta_j = \left[\langle \mathrm{AMSE}_j(x^{d_0}, x^{d_0 + \tau}) \rangle_{\text{year}} \right]^{-1},
\end{equation}

\noindent where $\langle ... \rangle_\mathrm{year}$ represents the time-average. Calculation of the per-variable AMSE contribution over a single year was found to be sufficient as the inter-year variance of $\beta_j$ was minimal. 

\subsubsection*{$k$-dependent weights}
\label{sec:k_dep_weights}

\noindent Atmospheric dynamics cause the power spectral density (PSD) of forecast fields to decay rapidly with increasing wavenumber $k$ (i.e., smaller scales), typically following power laws with scale-dependent exponents. At synoptic and subsynoptic scales ($\SI{5000}{km}$–$\SI{1000}{km}$), dynamics such as Rossby waves and jet streams dominate, producing a $k^{-3}$ scaling~\citep{Baer1972}. At smaller scales ($\SI{500}{km}$–$\SI{5}{km}$), turbulence becomes important, yielding shallower spectra of $\sim k^{-5/3}$ or $k^{-1}$~\citep{Gage1979}. Consequently, in the AMSE calculation, large-scale differences (small $k$) dominate because the PSD is orders of magnitude larger, while small-scale errors ($k \gtrsim 50$) are not penalised as effectively. This weighting is often desirable, since accurate representation of large-scale dynamics is critical, but it also limits sensitivity to high-$k$ errors.

In an attempt to overcome this, a $k$-dependent weighting was introduced to the MSH loss function: 

\begin{equation}
\gamma_k = \mathrm{max} \left[ \mathcal{N}_k \cdot k^{\sqrt{3}}, 1.0 \right],
\end{equation}

\noindent where $\mathcal{N}_k$ is a normalisation factor that preserves the magnitude of the AMSE calculation for a given variable and is used to maintain stable gradient norms during training. Use of this $k$-dependent weighting scheme addresses the imbalance in contributions between spectral errors over the range of $k$, while still maintaining the largest contribution at synoptic scales. 


\subsubsection{MSH loss calculation}
\noindent The loss metric used when training the FastNet model with the MSH loss function was therefore: 

\begin{align}
    \mathcal{L}_\mathrm{MSH} = \frac{1}{| D_\mathrm{batch}|} & \sum_{d_0 \in D_\mathrm{batch}} \frac{1}{T_\mathrm{train}} \sum_{\tau \in 1:T_\mathrm{train}} \sum_{j \in J} s_j w_j \beta_j \notag \\ & \times \sum_k \gamma_k \mathrm{AMSE}_{j,k}(\hat{x}^{d_0 + \tau}_j, x^{d_0 + \tau}_j),
\end{align}

\noindent where, in addition to the symbols already defined in Section \ref{sec:fastnet_model},

\begin{itemize}
    \item $\beta_j$ is the per-variable-level AMSE correction factor,
    \item $\gamma_k$ represents the $k$-dependent weights,
\end{itemize}

\subsubsection{MSH fine-tuning schedule}
\label{sec:MSH-FT-sched}

\begin{figure*}[!t]
\noindent\includegraphics[width=\textwidth]{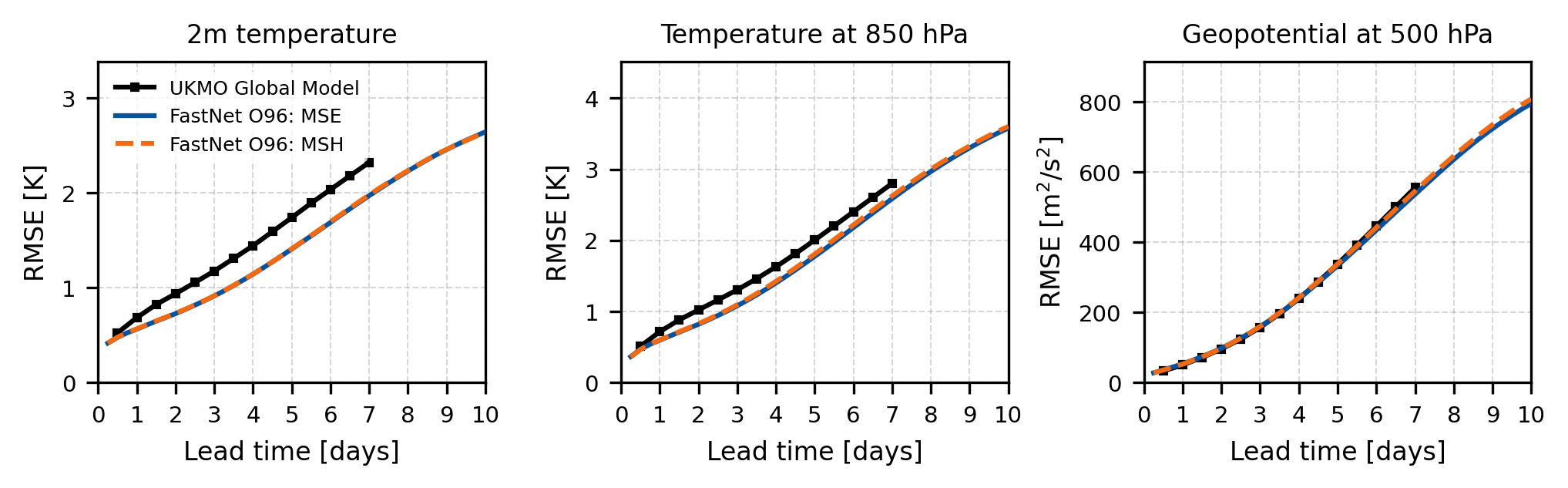}
\caption{RMSE values as a function of lead time for the Met Office Deterministic Global Model (black), MSE-fine-tuned FastNet O96 model (blue), and MSH-fine-tuned FastNet O96 model (orange, dashed). All model outputs are regridded to \ang{1.5}, compared to ERA5 reanalysis datasets and evaluated in 2022.}
\label{fig:RMSE-MSH-vs-MSE}
\end{figure*}

\noindent  The MSH loss function is used as an \textit{additional} fine-tuning step following MSE fine-tuning of the model, and uses the schedule defined in Table~\ref{tab:msh_sched}. Fine-tuning used the same ERA5 training dataset ($1980 - 2020$ inclusive), a batch size of 8 and a fixed learning rate was used for all steps. It was empirically found that this schedule provided a good balance between combatting the ``blurring" introduced by training using the MSE loss function while minimising the detrimental effect to the RMSE scores, thus maintaining good model accuracy at long lead times while simultaneously improving the physical consistency of the model predictions.

\begin{table}[h]
    \centering
    \begin{tabular}{ccc}
        \toprule
        Lead times & $\#$ steps & Learning rate \\
        \midrule
        1  & \SI{36525}{} & \SI{1e-5}{} \\
        2  & \SI{7305}{} & \SI{1e-7}{} \\
        4  & \SI{7305}{} & \SI{1e-7}{} \\
        8  & \SI{7305}{} & \SI{1e-7}{} \\
        12 & \SI{7305}{} & \SI{1e-7}{} \\
        \bottomrule
    \end{tabular}
    \caption{Fine-tuning schedule used to further train the MSE-fine-tuned FastNet model using the MSH loss function.}
    \label{tab:msh_sched}
\end{table}

\subsubsection{MSH Results}
\label{sec:MSH_results}

\begin{figure*}[t]
\noindent\includegraphics[width=\textwidth]{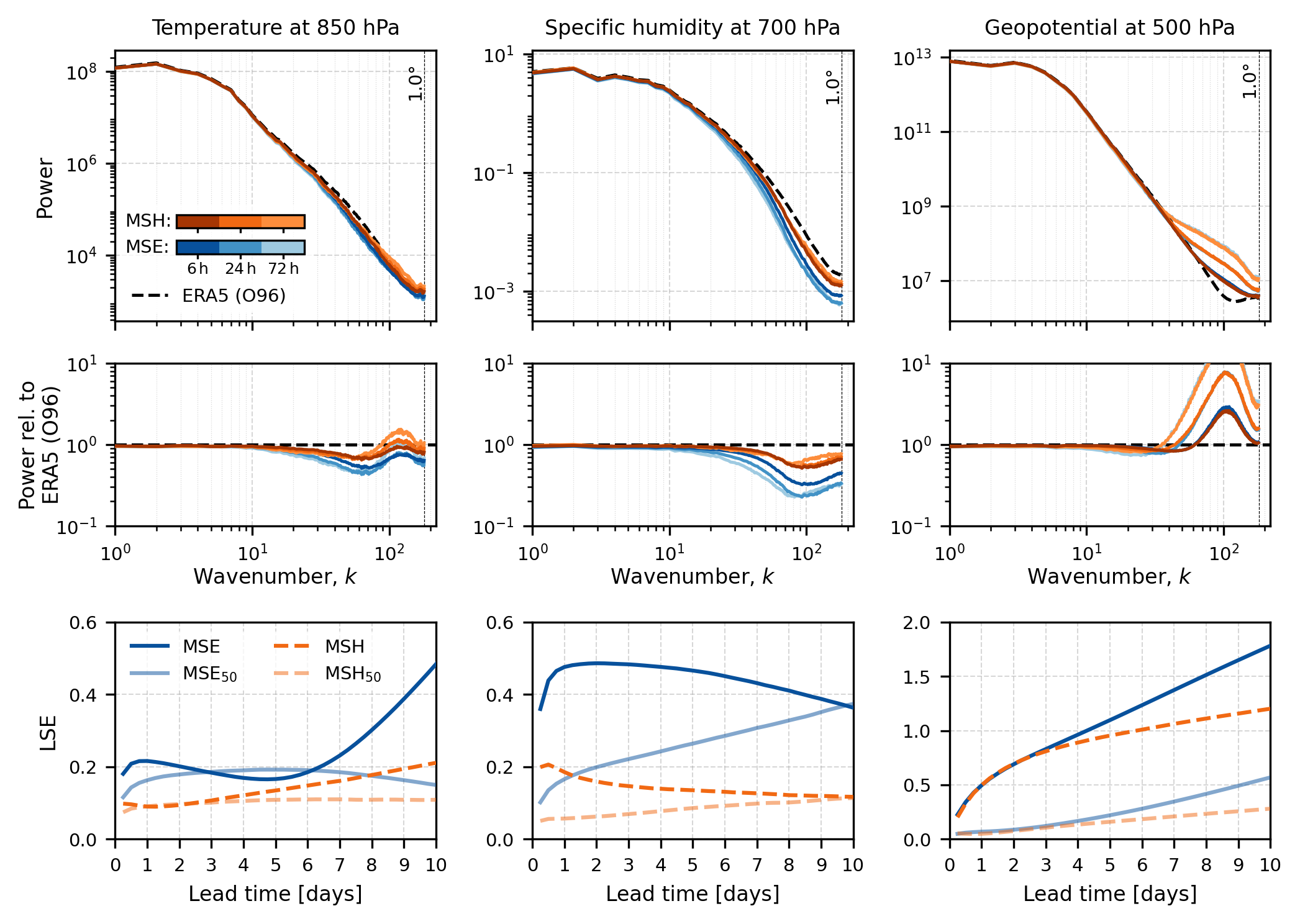}
\caption{Effect of loss function on power spectra. Top: Power spectrum of forecast predictions for the MSE fine-tuned (blue) and MSH fine-tuned (orange) FastNet O96 models at three different lead times: 6, 24 and 72 hours. The corresponding power spectra of the ERA5 (O96) reanalysis dataset is represented by the black, dashed line. All datasets are regridded to a regular \ang{1.0} lat/lon grid before the power spectra are calculated. Middle: the ratio of the power of the predictions compared to ERA5 O96. Bottom: logarithmic spectral error (LSE, see text for definition) as a function of lead time where the sum is performed over all $k$ (solid), and truncated at $k = 50$ (translucent). All results are evaluated in 2022.}
\label{fig:power-MSH-vs-MSE}
\end{figure*}

\noindent The effect of MSH fine-tuning on the performance of the O96 FastNet model are shown in Fig.~\ref{fig:RMSE-MSH-vs-MSE}. Both MSE- and MSH-fine-tuned FastNet models outperform the Global Model across all lead times, except for geopotential at \SI{500}{hPa} where performance is initially better for the Global Model but is surpassed by both FastNet models at 6 days. MSH fine-tuning yields RMSE scores very close to, though slightly higher than, MSE fine-tuning alone (differences of only a few percent), reflecting its focus on reducing spectral error rather than optimising MSE and thereby mitigating the “blurring” effect.

The main advantage of use of the MSH loss is evident in the power spectra (Fig.~\ref{fig:power-MSH-vs-MSE}). Both models match ERA5 on large scales (small $k$), but the MSE-trained model shows the typical “blurring” for $k \gtrsim 10$: power decays more rapidly than for the reanalysis as finer features are smoothed to minimise MSE, an effect that worsens with lead time. By contrast, the MSH model preserves power over a much broader range of $k$ and shows less lead-time dependence, demonstrating its ability to limit smoothing. In order to quantify the changes to the power spectra introduced via use of the MSH loss function, we calculate their logarithmic spectral error (LSE):

\begin{equation}
    \mathrm{LSE} = \sqrt{\frac{1}{K} \sum_{k=0}^K\left[\log_{10}\mathrm{PSD}_k (\hat{x}) -  \log_{10}\mathrm{PSD}_k (x)\right]^2},
\label{eq:LSE}
\end{equation}

\noindent the root mean square error between the log-transformed predictions and ground-truth power spectral densities. This error formulation reduces the dynamic range of the spectrum and weights all frequencies equally in $\mathrm{log}$-space, allowing errors at small scales (large $k$) to still contribute significantly despite their reduced magnitude. The results for the MSE- and MSH-fine-tuned models are shown in the bottom panel of Fig.~\ref{fig:power-MSH-vs-MSE}, and demonstrate significant improvement in LSE as a function of lead time for the MSH-tuned model thanks to its increased power in the region $10 \lesssim k \lesssim 100$. 

Unlike point-wise RMSE whose values typically grow with increasing forecast lead time, the LSE values for \texttt{T850} and \texttt{q700} can be seen to improve with increasing lead time as the power spectra evolve. This is due to a feature evident in the power spectra of both the MSE- and MSH-fine-tuned models: an unphysical up-tick in the power near the smallest scales ($k \sim 100$) which represents a significant deviation from the reanalysis data. It is most obvious in the predictions of the geopotential, but is generally present across all forecast variables. It can also be seen that the magnitude of this up-tick grows with increasing lead time, working against the blurring effect that reduces power at small scales and, in some cases, artificially improving the LSE scores of the model output as its PSD moves closer to the ground truth. Truncating the sum over $k$ in Eq.~(\ref{eq:LSE}) at $k = 50$ gives the spectral error only at the largest scales, minimising any contribution from the unphysical up-tick, as demonstrated by the translucent curves in the lower panel of Fig.~\ref{fig:power-MSH-vs-MSE}. This better illustrates the gradual worsening of the spectra with increasing lead time due to blurring within the MSE-fine-tuned model, and again highlights the significant improvement when using the MSH loss function.

\begin{figure*}[t]
\noindent\includegraphics[width=\textwidth]{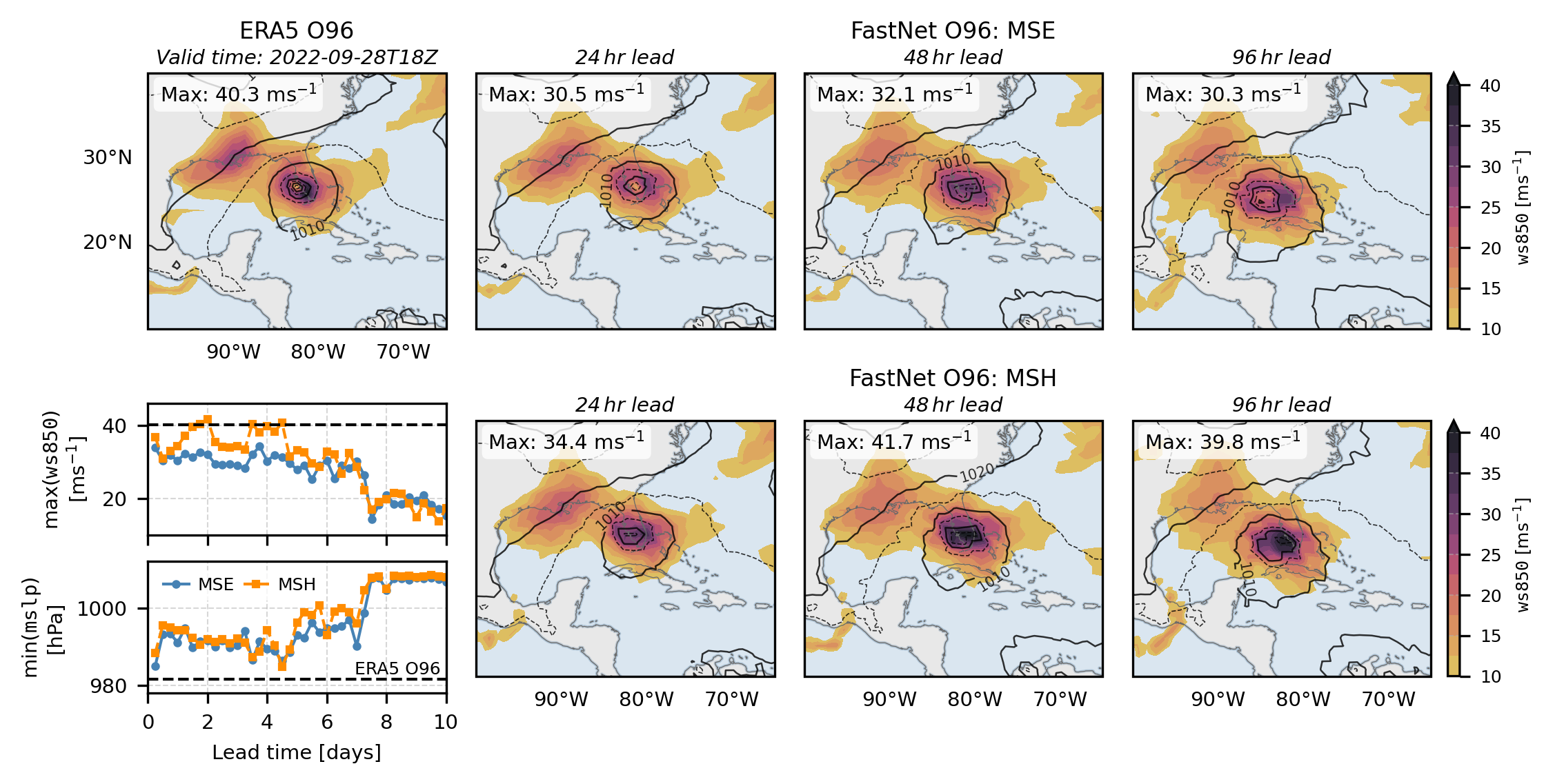}
\caption{Wind speed at \SI{850}{hPa} (colours) and MSLP structure (contours) around the peak of Hurricane Ian. Top left: ERA5 reanalysis, valid at 18 UTC 28 September 2022. Top row: predictions from MSE-fine-tuned FastNet O96 model at 24, 48 and 96 hour lead times. Bottom row: Equivalent for MSH-fine-tuned model. Bottom left: Maximum forecast wind speed at \SI{850}{hPa} (\texttt{ws850}) and minimum mean sea level pressure (\texttt{mslp}) within the plotted region as a function of lead time for the MSE- (blue) and MSH-fine-tuned (orange) models. The horizontal dashed line represents the equivalent values from the ERA5 reanalysis data.}
\label{fig:MSH-vs-MSE_hurricane}
\end{figure*}

However, use of the MSH loss function does not remove or diminish the unphysical up-tick within the spectra, indicating that its presence may not be loss-function-dependent but that its source is inherent to the model architecture. The wavenumber at which this increase in power occurs corresponds to an angular resolution of approximately \ang{2.0} --- the resolution of the internal processor mesh of the FastNet O96 model. We discuss this feature, and a method for overcoming it, further in Section~\ref{sec:horz_grads}.

\subsubsection*{Effect on predictions of extremes}

\noindent Use of the MSH loss function to further fine-tune the model can help to improve predictions of extreme events. Figure~\ref{fig:MSH-vs-MSE_hurricane} shows predictions produced by the FastNet O96 models compared to the ERA5 reanalysis dataset for Hurricane Ian, an out of sample category 5 hurricane that struck western Cuba, Southern US and the Carolinas in late September 2022, causing widespread damage and significant loss of life. The predictions are valid at 18 UTC, 28 September 2022, near the intensity peak of the hurricane as it made landfall on south-western Florida. The MSE fine-tuned model significantly underestimates the peak wind speeds and produces predictions that are far smoother, particularly around the central depression. In contrast, further fine-tuning with the MSH loss function recovers the structure of the hurricane at longer lead times despite the relatively low resolution of the model (approx. equivalent to \ang{1.0} lat/lon, $\sim \SI{110}{km}$), with peak wind speeds that are more accurate from a lead time of approximately five days. Despite the improvements in structure and wind speed predictions for the MSH fine-tuned model, both models underestimate the magnitude of the drop in pressure at the centre of the hurricane, likely constrained by their coarse resolution.

\begin{figure}[t]
\noindent\includegraphics[width=0.5\textwidth]{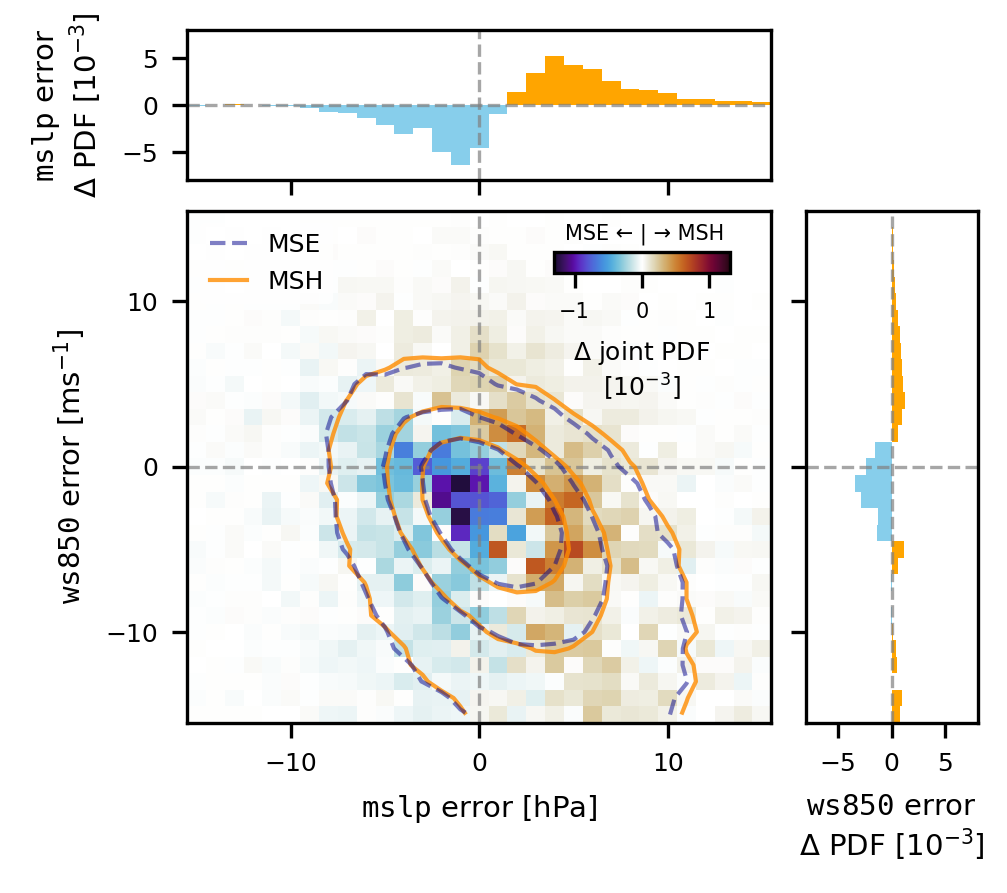}
\caption{Joint-PDFs of errors in FastNet model predictions (contours) of \texttt{mslp} and \texttt{ws850} for wind speed events that exceed the 99.5$^\mathrm{th}$ percentile ($\texttt{ws850} > \SI{30}{ms^{-1}}$) in 2022 at lead times of 72 hours. The contours contain $[50, 75, 95]\%$ of the probability mass for each model --- MSE fine-tuned (blue, dashed) and MSH fine-tuned (orange). The difference between the joint-PDFs is represented by the coloured background, with orange (blue) representing an increased frequency of error magnitudes for the MSH (MSE) fine-tuned model. Top: difference in marginal error distributions for \texttt{mslp} predictions. Right: as in top, for \texttt{ws850} predictions.}
\label{fig:MSH-vs-MSE_errorPDFs}
\end{figure}

A more detailed analysis of the two models' predictions for intense storms can be found in Fig.~\ref{fig:MSH-vs-MSE_errorPDFs}, which shows the joint probability distribution functions (PDFs) of their errors in predictions of \texttt{ws850} and \texttt{mslp} at 72 hour lead times. Errors in predictions of these variables can provide insight into the ability of the models to capture storm intensity and characterize the strength and structure of extreme weather systems in a physically consistent way. The most intense storms were selected from the ERA5 dataset by considering only events with wind speed at \SI{850}{hPa} that exceeded the 99.5$^\mathrm{th}$ percentile over the entirety of 2022 --- a threshold that corresponded to $\texttt{ws850} > \SI{30}{ms^{-1}}$. The joint PDFs are represented by the contours and contain $[50, 75, 95]\%$ of the total probability mass for each model. They appear very similar for both models and indicate that, in general, both models provide reasonably accurate predictions of the \texttt{mslp} in the vicinity of intense storms but have a tendency to under-predict the wind intensity. Comparing the PDFs directly --- represented by the coloured background of Fig.~\ref{fig:MSH-vs-MSE_errorPDFs} --- illustrates subtle, but important, differences in the physical consistency of the two models. Blue regions indicate areas of the error space where the MSE fine-tuned model's predictions are more frequent, suggesting that this model tends to over-predict storm central pressure deepening (i.e., more negative MSLP errors) while simultaneously under-predicting wind intensity—placing it in the lower-left quadrant of Fig.~\ref{fig:MSH-vs-MSE_errorPDFs}. This pattern is physically inconsistent, as deeper central pressure typically corresponds to stronger winds due to an enhanced pressure gradient. In contrast, the MSH fine-tuned model tends to under-predict the magnitude of pressure deepening (more positive MSLP errors) but does so in tandem with weaker wind predictions, indicating a more physically consistent relationship between pressure and wind fields. The right panel of Fig.~\ref{fig:MSH-vs-MSE_errorPDFs} shows that the MSH fine-tuned model more frequently predicts stronger wind speeds compared to the MSE fine-tuned model, consistent with the results in Fig.~\ref{fig:MSH-vs-MSE_hurricane}, and also suggests a tendency for the MSH model to overestimate wind intensity during extreme events.

\subsection{Inclusion of horizontal gradients within the loss}
\label{sec:horz_grads}
\noindent It was noted in Section \ref{sec:MSH_results} that non-physical patterns begin to appear in the output of data-driven weather models, particularly at long lead times~\citep{keisler_forecasting_2022}. The exact nature of these patterns depends on the underlying architecture of the data-driven model~\citep{Ellis2025}, but they can lead to visible artefacts in the forecast fields and corresponding power spectra --- artefacts that are particularly prevalent in the horizontal gradients of the fields. Given that these artefacts are not present in the training data and are non-physical, their presence may limit the usefulness of these fields and their use in derived diagnostics, making it important to find methods to remove them from forecast predictions of data-driven models.

One approach explored within training of the FastNet model was to include calculations of the horizontal gradients of forecast and target fields within the loss at each training step. This encourages the model to not only produce accurate predictions of the fields themselves, but also penalises predictions of fields whose horizontal gradients contain non-physical artefacts. We note here that the model is \textit{not} predicting the horizontal gradients of fields directly, it is only being rewarded/penalised according to the accuracy of the derived fields in combination with the raw fields.

\subsubsection{Practical implementation}

\noindent In order to effectively train on the horizontal gradients of the forecast fields alongside the raw fields, a method for efficiently calculating the zonal and meridional derivatives of a given batch of forecast (or target) variables is necessary. This is achieved by pre-computing derivative operators $D_u$, such that:

\begin{equation}
    \frac{\partial f}{\partial u} \approx D_u f, \; u \in \{x, y\},
\end{equation}

\noindent and therefore calculation of the derivatives of the forecast fields requires only two matrix multiplication operations --- one for each of the zonal and meridional derivatives. The derived fields are then appended to the feature dimension of the batch and the loss calculation continues as before. This approach means that either the MSE or MSH loss function can be used. The per-variable-level weights associated with the loss from the derived fields can be tuned as necessary, but here we chose that they should match the per-variable weighting scheme used throughout.

\begin{figure*}[t]
\noindent\includegraphics[width=\textwidth]{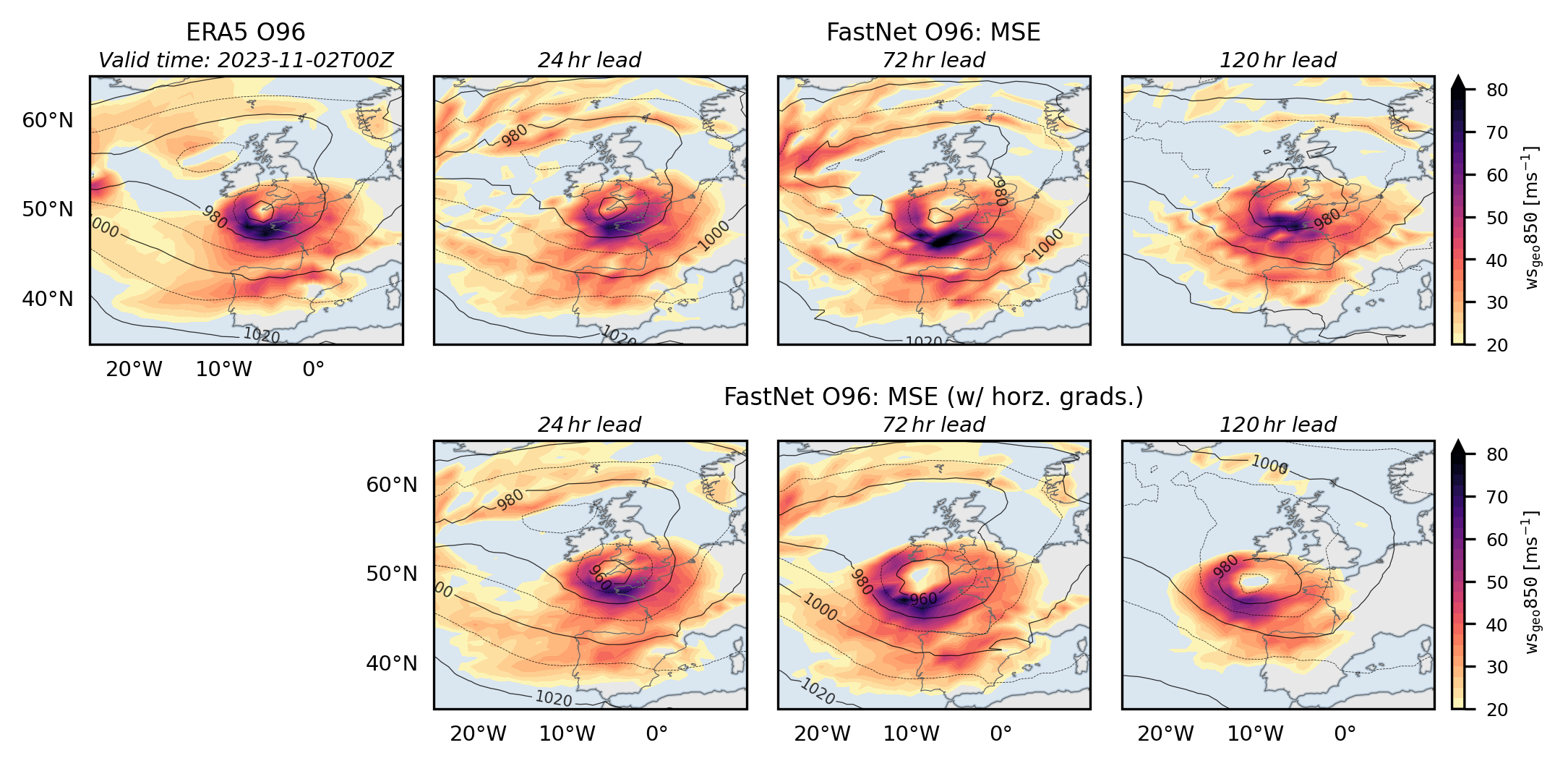}
\caption{Artifical windspeed features in ML predictions. Geostrophic wind speed at \SI{850}{hPa} (colours) and MSLP structure (contours) around the peak of Storm Ciaran from ERA5 reanalysis and FastNet O96 MSE fine-tuned. All plots correspond to valid times of 00 UTC 2 November 2023. Top: FastNet O96 MSE fine-tuned model at lead times of 24 (left), 72 (middle) and 120 (right) hour lead times. Bottom: As top, but for a FastNet O96 MSE model trained with horizontal gradients included in the loss calculation.}
\label{fig:O96-artefacts}
\end{figure*}

The reduced Gaussian grid upon which FastNet operates contains fixed latitudinal points but variable longitudinal spacing. Standard 4th-order central difference weights can therefore be used for the zonal ($\partial_x$) derivative, as neighbours lie along the same latitude. In contrast, north–south neighbours required for the meridional ($\partial_y$) derivative do not necessarily align longitudinally, so finite difference weights must be recalculated. For each grid point, the nearest neighbours in the two adjacent latitude bands are identified, and the projection of the connecting vector onto the northward direction is computed, accounting for Earth’s curvature. Fornberg’s method~\citep{Fornberg2002} is then applied to obtain the finite difference weights. This also enables forward/backward differences near the poles where central differences are not possible. All derived fields are normalised to zero mean and unit variance so their contribution to the loss is comparable to that of the raw fields (prior to per-variable-level weighting).

This gradient computation has negligible effect on training rate under both MSE and MSH losses. Memory usage increases slightly --- $+0.6\%$ on average for MSE up to 6 lead times, $+1.2\%$ for MSH, on the O96 grid --- but not enough to alter the training schedule. Given this minimal cost, the model was pre-trained with horizontal gradients in the loss function, rather than adding them only in fine-tuning. This approach best suppressed artefacts in the predictions while limiting RMSE impact. By contrast, introducing gradients solely at fine-tuning led to rapid RMSE degradation, while artefact reduction appeared only after extensive training ($\sim30\%$ of the pre-training schedule).

\subsubsection{Gradient loss results}

\noindent Given that model artefacts are most visible in the gradients of the predictions generated by data-driven models, it is instructive to consider derived fields such as the \textit{geostrophic} wind --- an idealised wind that arises from the balance between the pressure gradient and Coriolis forces. The $u$- and $v$-components of the geostrophic wind can be calculated from derivatives of the geopotential, $\Phi$, as:

\begin{equation}
    \begin{pmatrix} u_\mathrm{geo} \\ v_\mathrm{geo} \end{pmatrix} = \frac{1}{f} \begin{pmatrix} -\partial_y \Phi \\ \partial_x \Phi \end{pmatrix},
\end{equation}

\noindent where $f = 2 \Omega \sin \phi$ is the Coriolis parameter, $\phi$ is the latitude and $\Omega = \SI{7.292e-5}{rad s^{-1}}$ is the Earth's angular velocity.

\cite{charlton-perez_ai_2024} studied the medium-range predictions provided by data-driven models in comparison to those produced by NWP models, focusing on their predictions of Storm Ciar\'an, an extratropical cyclone ``bomb'' that struck north-western Europe at the beginning of November 2023. Within the study it was shown that non-physical structures (artefacts) were clearly visible in the calculated geostrophic wind speed for the data-driven models considered and so we have chosen to use the same event to study the effect of including horizontal gradients of forecast fields within the loss calculation. We note that the FastNet models considered in this study use data defined on an O96 grid (\ang{1.0} grid spacing) and as such have reduced resolution compared to the results discussed in \cite{charlton-perez_ai_2024} which were all evaluated at \ang{0.25}. This reduces the peak wind speed intensity and the magnitude of the pressure deepening compared to the higher resolution data, however model artefacts are still present.

\begin{figure*}[t]
\centering
\noindent\includegraphics[width=0.6\textwidth]{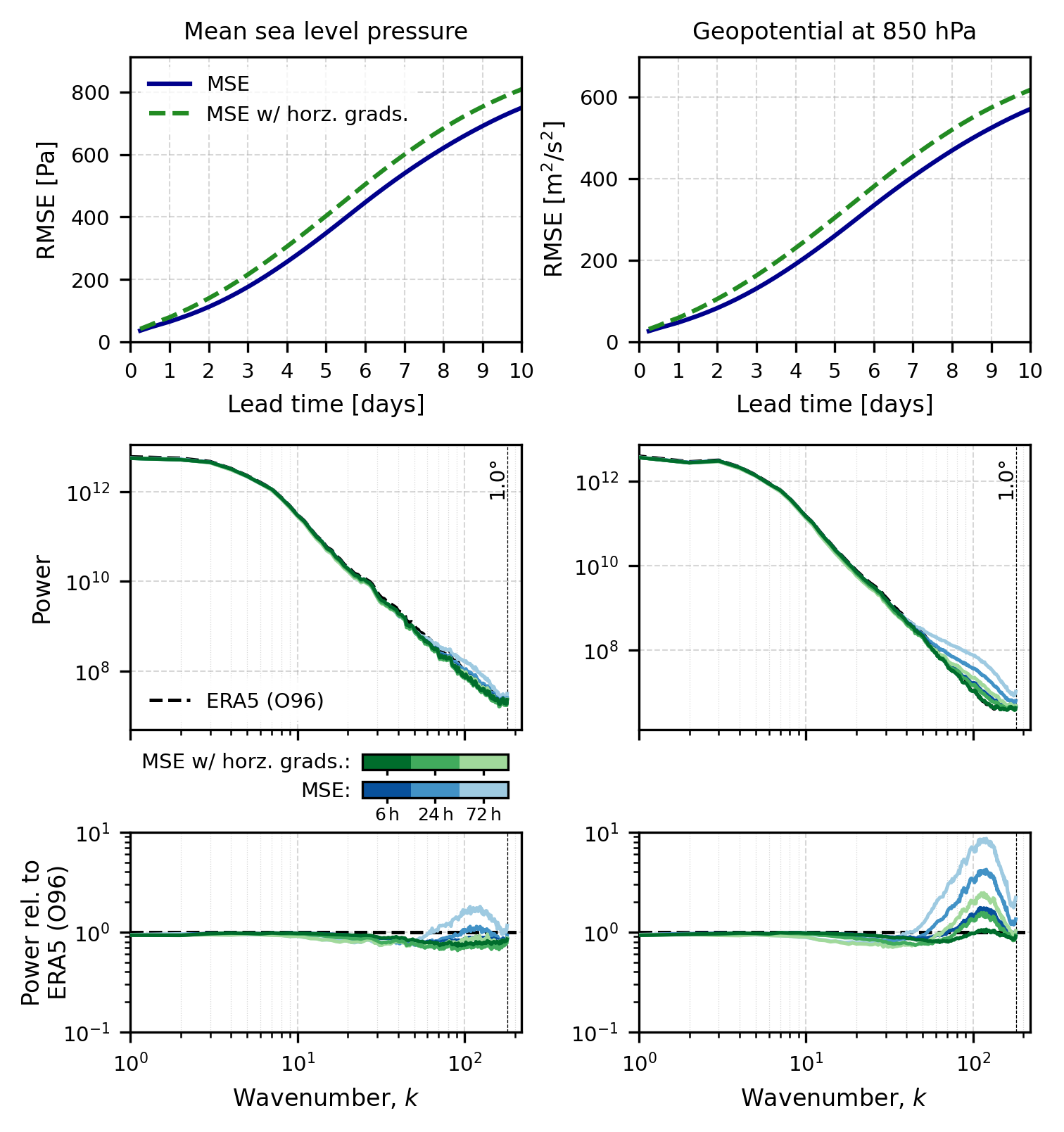}
\caption{Top: RMSE scores for the baseline MSE fine-tuned FastNet O96 model (blue) and an equivalent trained with horizontal gradients included in the loss calculation (green, dashed). The corresponding power spectra are shown in the middle panels, with the power relative to the ERA5 reanalysis dataset on the bottom panels. All data are evaluated over 2022.}
\label{fig:O96-artefacts-RMSE-PS}
\end{figure*}

Figure~\ref{fig:O96-artefacts} shows a comparison of the calculated geostrophic wind speed, $ws_\mathrm{geo} = (u_\mathrm{geo}^2 + v_\mathrm{geo}^2)^{1/2}$, at \SI{850}{hPa} and mean surface level pressure (MSLP) at 00 UTC, 2 November 2023, around Storm Ciar\'an's peak intensity as it made landfall on the Cornish coast. The predictions of two versions of the O96 FastNet model are shown alongside the corresponding ERA5 reanalysis: the top row shows the baseline MSE fine-tuned FastNet model, while the bottom row shows an equivalent model trained with horizontal gradients included in the loss calculations. Non-physical, honeycomb structures are clearly visible in the geostrophic wind maps for the baseline model (Fig.~\ref{fig:O96-artefacts}, top row). The structures can be seen at all lead times, but their visibility grows with lead time. This is mirrored in the MSLP contours whose structure is unrealistically sharp and noisy, particularly at long lead times. The pattern of the honeycomb structures matches the positions of the internal processor mesh, confirming the source of the increased power spectral density observed at this frequency (cf. Fig~\ref{fig:power-MSH-vs-MSE}). In contrast, when horizontal gradients are included in the loss calculation (Fig.~\ref{fig:O96-artefacts}, bottom row) the geostrophic wind speed field is smoother indicating improved physical consistency of the model predictions. The honeycomb pattern is no longer visible, even at \SI{120}{hour} lead times, and the MSLP contours are far smoother, indicating that this approach has successfully reduced the model artefacts.

We note however that inclusion of the horizontal gradients in the loss calculation has a worsening effect on the RMSE scores of the model. This can be visualised in Fig.~\ref{fig:O96-artefacts} where, at long lead times, the storm centre appears to be shifted significantly to the west of its actual location. Missing the location of the storm results in the model being penalised twice, significantly worsening its RMSE scores as shown in Fig.~\ref{fig:O96-artefacts-RMSE-PS}. The baseline model outperforms that trained with horizontal gradients included, although we note that the gradient-trained model still maintains improved RMSE compared to the Met Office Deterministic Global Model. In contrast, the power spectra show significant improvement when the horizontal gradients are included in the loss calculations. The increase in power around wavenumber $k \sim 100$ observed in both the MSE- and MSH-trained versions of the FastNet model is much reduced --- and effectively eliminated for most variables, such as the MSLP --- however a small increase in power remains in the geopotential fields.

It is worth noting that in the spectral domain, horizontal derivatives can be calculated by multiplying the spectral coefficients by a term proportional to the horizontal wavenumber $k$. Inclusion of horizontal gradients in the loss function therefore has the effect of upweighting the higher wavenumbers with linear dependence on $k$, and so may counteract some of the weight optimisations described in Section \ref{sec:k_dep_weights}. We further explore the addition of horizontal derivatives to the MSH loss function during the combined training regime in Section~\ref{sec:consolidated_results}.

\subsection{Alternative representations of wind input variables}
In NWP models, wind is typically represented by its horizontal vector components: the zonal component $u$ (east-west flow) and the meridional component $v$ (north-south flow). However, this representation permits neural networks to learn to compensate for directional errors in the training data by reducing the wind speed, which, for small reductions in magnitude, can reduce the total RMSE for the individual wind components. This tendency for dampened wind speeds to be rewarded by a reduced RMSE is not unique to MLWP models, but is also a known characteristic of NWP models. 

For many applications - particularly those involving high-impact, extreme wind events - errors in wind speed are more important than errors in wind direction. In addition, speed and direction have different error characteristics in the training data, and so should be weighted differently in the loss function for optimal training. This motivates an alternative representation of wind in terms of separate speed and direction components.

Only the loss function is modified in this approach; inputs to the model remain as normalised individual wind components.
Wind speed, $s$, is defined as
\begin{equation}
s = \sqrt{u^2 + v^2}.
\end{equation}
\noindent 
Directional information is captured through unit vectors of $u$ and $v$, computed as:

\begin{equation}
\boldsymbol{\overrightarrow{d}} = \left ( \frac{u}{s}i + \frac{v}{s}j \right ).
\end{equation}

This encodes wind direction as coordinates on the unit circle, providing a representation that is both computationally efficient and numerically stable compared to inputting wind direction directly. This is particularly beneficial in machine learning workflows, where backpropagation through inverse trigonometric functions (such as $\arctan$) can lead to increased gradient variance and discontinuities in derivatives near quadrant boundaries. 

\subsubsection{Practical implementation}
The $u$ and $v$ input variables used in FastNet consist of one 10 metre height variable and 13 pressure-level variables. In this implementation, these variables are replaced in the loss calculations by their corresponding normalised $d_i$ and $d_j$ variables, with $s$ provided in addition. The number of wind input variables in the loss function is therefore increased by 14. The weighting scheme applied to $u$ and $v$ is retained for $d_i$ and $d_j$, while the additional $s$ variables are assigned weights equal to those of $u$, but scaled by a factor of five to prioritise wind speed accuracy during training.

\subsubsection{Results}
\begin{figure*}[h]
\noindent\includegraphics[width=\textwidth]{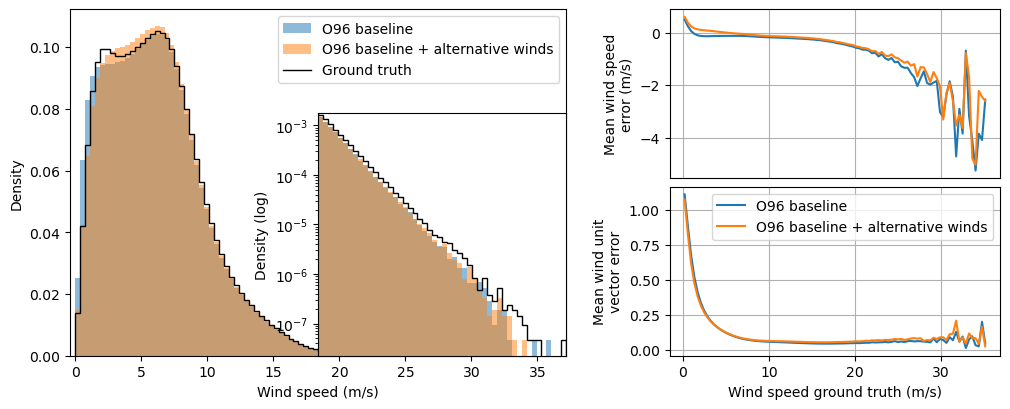}
\caption{Comparisons between the model trained with the alternative wind representation and the baseline model, evaluated against the ground truth for 10 m wind speed at a 6-hour lead time over 2022. Left: Probability density functions (PDFs) of the forecast and observed wind speeds, shown on a linear scale (main panel) and a logarithmic scale (inset). Right: Error metrics plotted against ground truth wind speed: top—mean wind speed error (bias); bottom—mean wind vector error (unit vectors).}
\label{fig:altwinds_dist}
\end{figure*}

The alternative representation of wind input variables has the desired effect of shifting the wind speed distribution toward higher values, and hence closer to the true distribution. Specifically, the baseline representation produces a notably higher density of low (near-zero) wind speeds compared to the alternative representation (Figure \ref{fig:altwinds_dist}, left). At low wind speeds, the distribution of the alternative representation aligns more closely with the ground truth data used to train the model. It also yields a greater proportion of higher wind speeds, broadly matching the ground truth. However, the log-scale reveals that the highest wind speeds at the tails of the distributions are under-predicted by both wind representations. This supports the idea that RMSE is minimised most effectively by reducing error in the more frequent lower wind speed regime.

The mean error (bias) in wind speed predictions for both models is mainly negative, indicating a tendency to under-predict wind speed, as expected (Figure \ref{fig:altwinds_dist}, right top). The alternative representation has a smaller negative bias, demonstrating an improvement over the baseline. At lower wind speeds, the alternative representation again outperforms the baseline in directional error (Figure \ref{fig:altwinds_dist}, right bottom). However at higher wind speeds the baseline has better directional accuracy, a direct result of our choice to prioritise wind speed accuracy in such conditions. These results demonstrate the ability to nudge the model predictions towards the most crucial aspects of their output by adjusting their representation in the loss function.

\begin{figure*}[h]
\noindent\includegraphics[width=\textwidth]{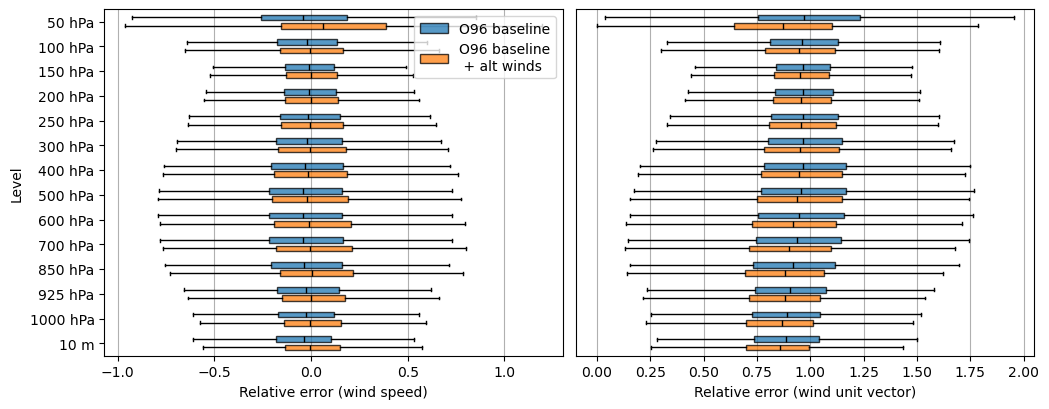}
\caption{Comparisons between the model trained with the alternative wind representation and the baseline model for relative wind speed error and relative wind unit vector error at a 72-hour lead time over 2022. Left: Boxplots showing the distribution of relative wind speed error for the alternative winds model (orange) and baseline model (blue) across all vertical levels included in training. Right: Boxplots of relative wind vector error for the same models and levels. Levels are ordered by ascending height/pressure, i.e. the lowest altitude at the bottom.}
\label{fig:altwinds_rel}
\end{figure*}

This trend persists across height/pressure levels and at a longer lead time, in this case 72 hours (Figure \ref{fig:altwinds_rel}). The alternative representation consistently yields improved relative errors than the baseline for both wind speed (Figure \ref{fig:altwinds_rel}, left) and wind unit vector (Figure \ref{fig:altwinds_rel}, right) predictions, with distributions (and medians) closer to zero across all levels in both cases. Notably, this improvement holds across the full range of wind speeds—including at the extremes of the distribution, demonstrating that decoupling speed and direction in the loss function enables the model to more effectively learn wind magnitude while preserving robust directional accuracy.

\section{FastNet v1.1: combined loss modifications}
\label{sec:consolidated_results}

\begin{figure*}[!t]
\noindent\includegraphics[width=0.75\textwidth]{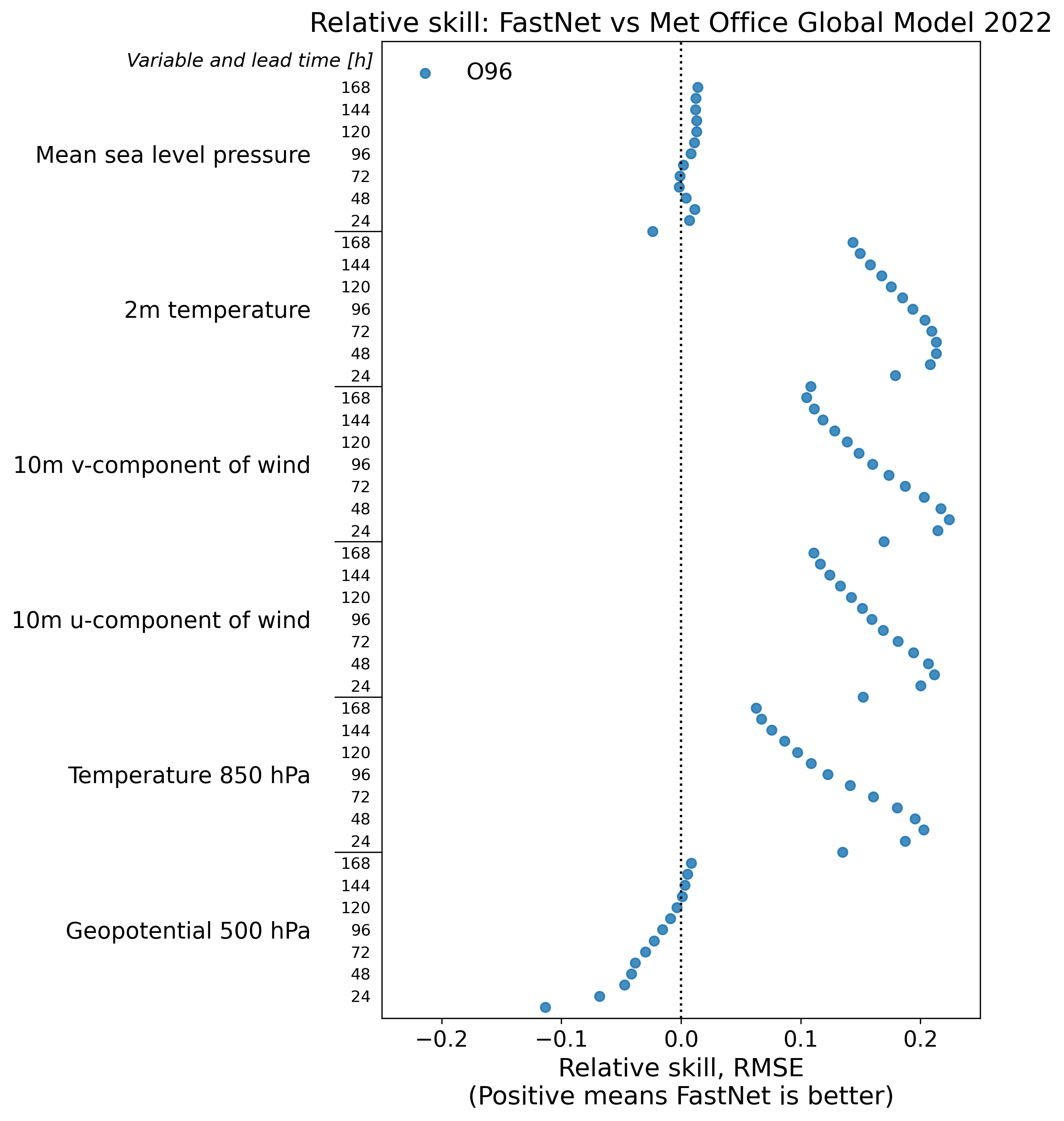}
\caption{Relative performance of FastNet v1.1 compared to the Met Office Deterministic Global Model (0.2 = 20\% improvement compared to the Global Model. All model outputs are regridded to a regular \ang{1.5} lat-lon grid and evaluated in 2022}
\label{fig:FNvsGM}
\end{figure*}

\begin{figure*}[!t]
\noindent\includegraphics[width=\textwidth]{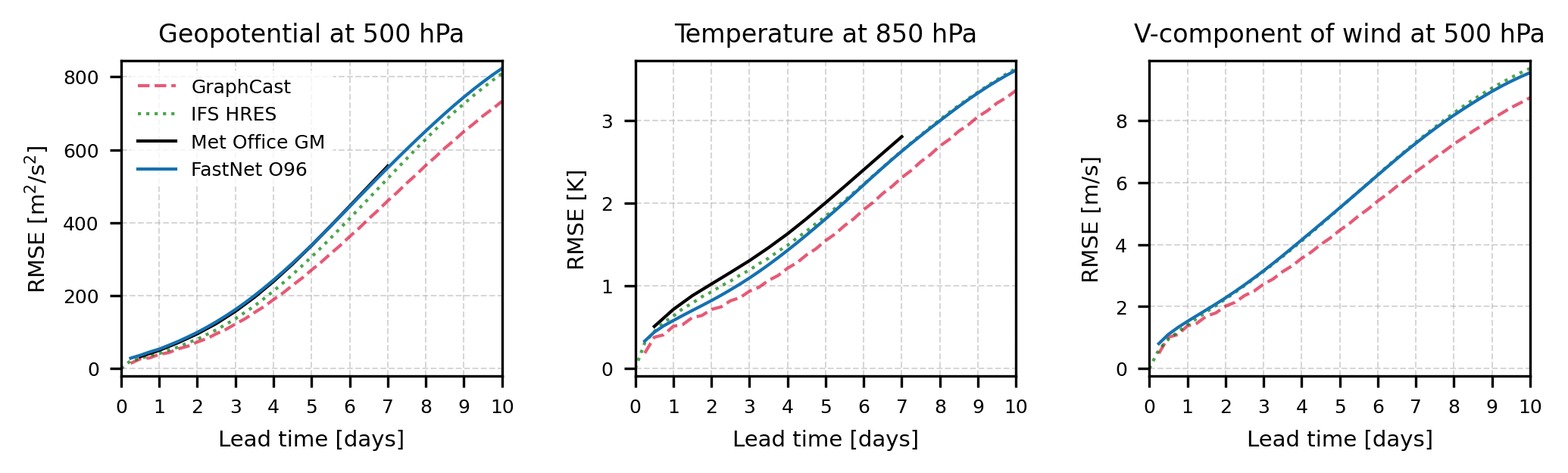}
\caption{RMSE values for upper-air variables as a function of forecast lead time for FastNet v1.1 compared to leading MLWP and NWP models. All model outputs are regridded to \ang{1.5}, compared to ERA5 reanalysis datasets and evaluated in 2022.}
\label{fig:RMSEvslead-upper}
\end{figure*}

\begin{figure*}[!t]
\noindent\includegraphics[width=\textwidth]{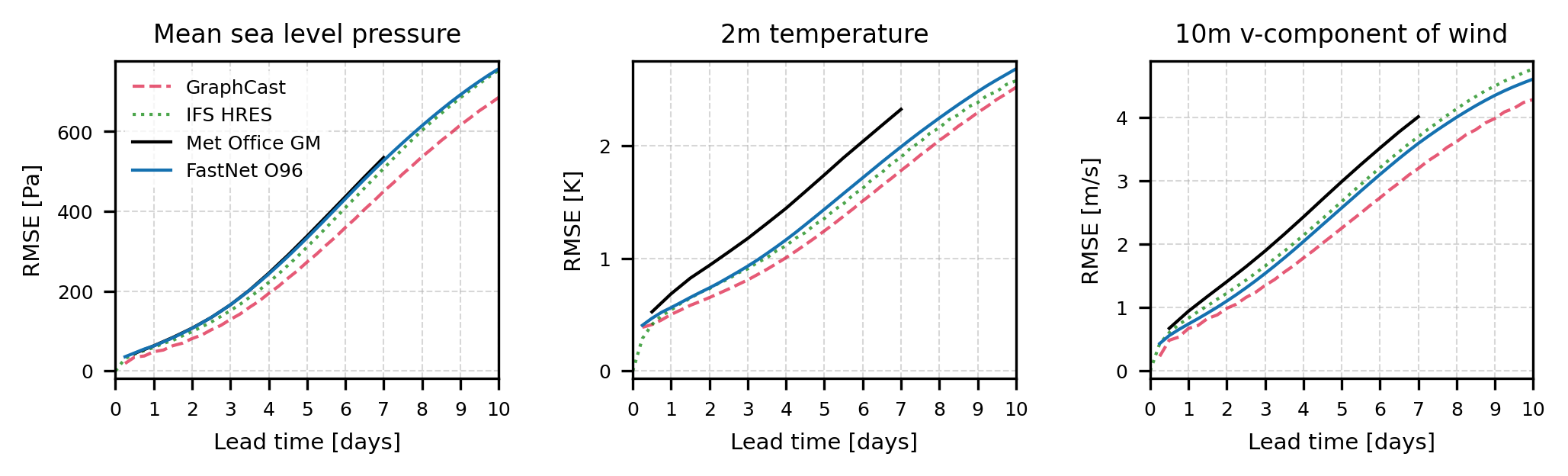}
\caption{RMSE values for surface variables as a function of lead time for FastNet v1.1 compared to leading MLWP and NWP models. All model outputs are regridded to \ang{1.5}, compared to ERA5 reanalysis datasets and evaluated in 2022.}
\label{fig:RMSEvslead-surface}
\end{figure*}

\begin{figure*}[h]
\noindent\includegraphics[width=\textwidth]{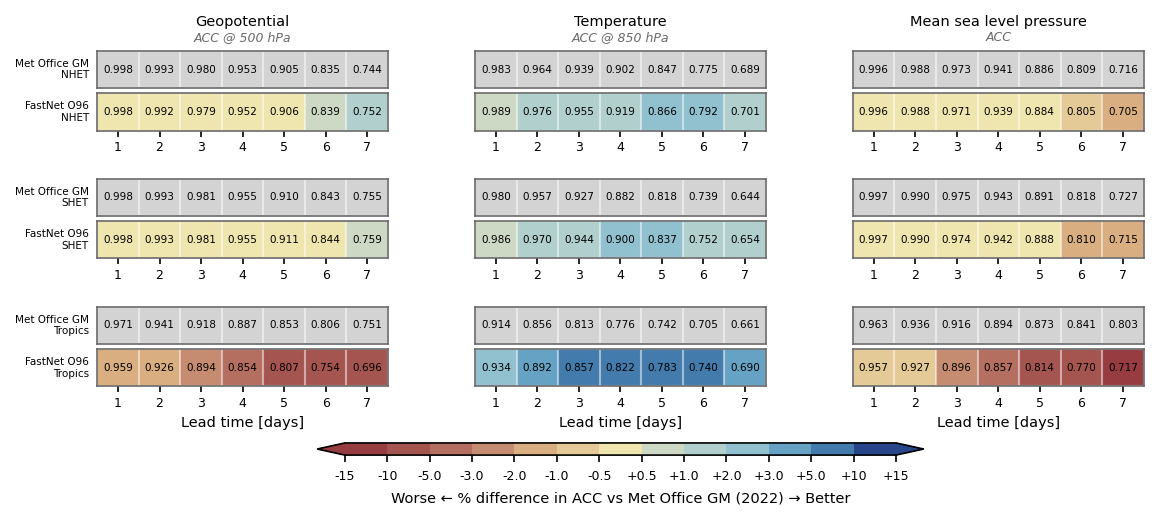}
\caption{ACC values for forecast variables up to, and including, lead times of 7 days; includes results for the Met Office Deterministic Global Model and the FastNet O96 v1.1 models The values of FastNet v1.1 is coloured according to its performance with blue (red) colours representing improved (decreased) ACC performance relative to the Met Office Global Model. Results for three regions on the globe are shown: northern/southern hemisphere extra-tropics (NHET/SHET) and the tropics. All model outputs are regridded to a regular \ang{1.5} lat-lon grid, compared to the equivalent ERA5 reanalysis dataset and evaluated in 2022.}
\label{fig:UKMO-consolidated-ACC-scorecard}
\end{figure*}

In this section we present FastNet v1.1, where all loss function modifications described in the preceding sections have been combined. Performance on standard metrics of RMSE and ACC are shown in Figures \ref{fig:FNvsGM}-\ref{fig:UKMO-consolidated-ACC-scorecard}. The RMSE performance of FastNet v1.1 exceeds that of the Met Office Deterministic Global Model for most variable-lead time combinations up to 7 days, with the exception of \texttt{Z500} at earlier lead times. We speculate that correcting the relatively poor performance of \texttt{Z500} is a matter of rebalancing the per-variable loss weights to increase the importance of geopotential, rather than anything more fundamental, but this will be explored in future work.

\begin{figure*}[t]
\noindent\includegraphics[width=\textwidth]{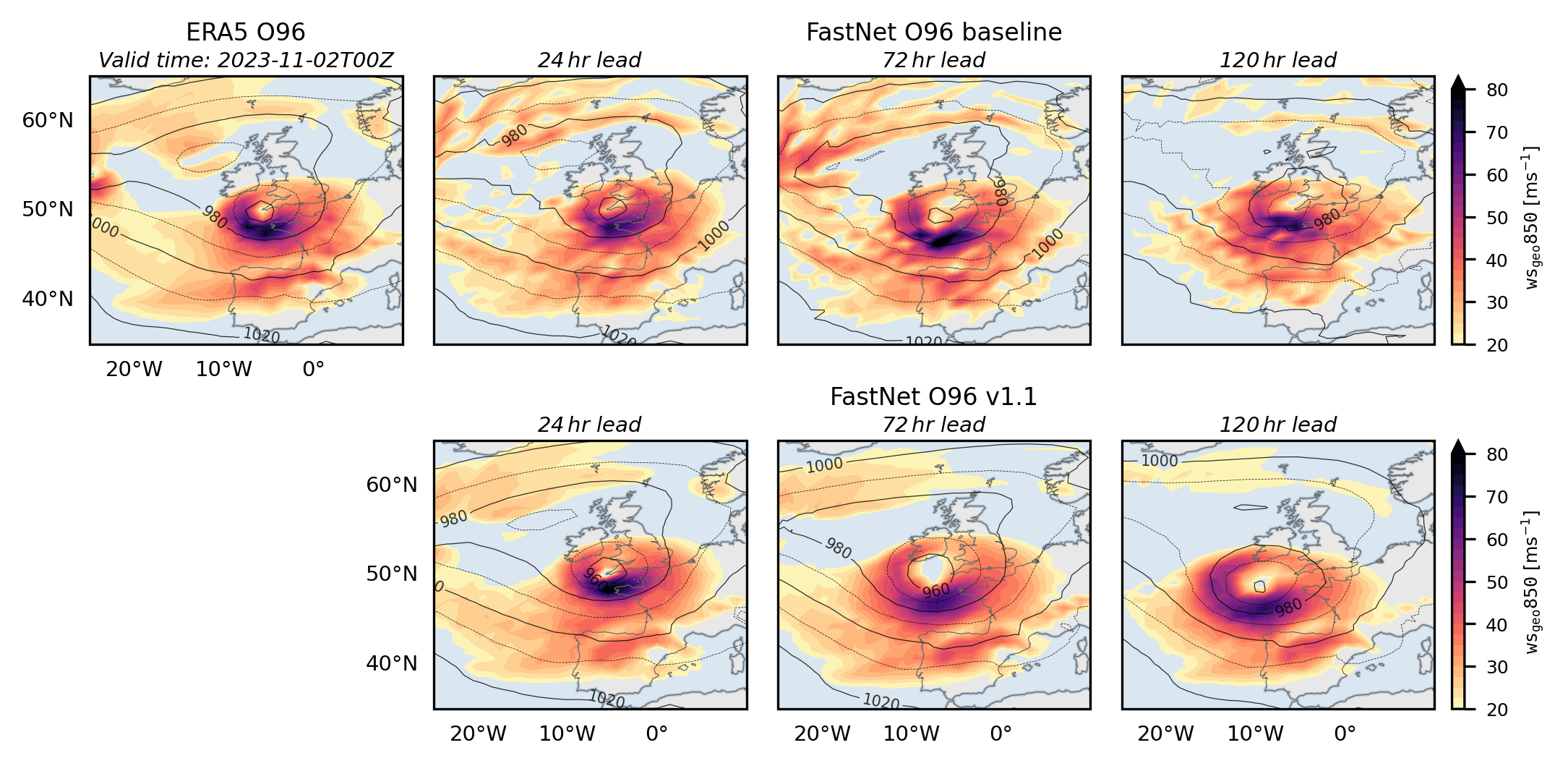}
\caption{As Figure \ref{fig:O96-artefacts} but for the FastNet v1.1 O96 model and O96 baseline. Artifical windspeed features in ML predictions. Geostrophic wind speed at \SI{850}{hPa} (colours) and MSLP structure (contours) around the peak of Storm Ciaran.}
\label{fig:fastnet_v1p1_ciaran}
\end{figure*}

\begin{figure*}[h]
\noindent\includegraphics[width=\textwidth]{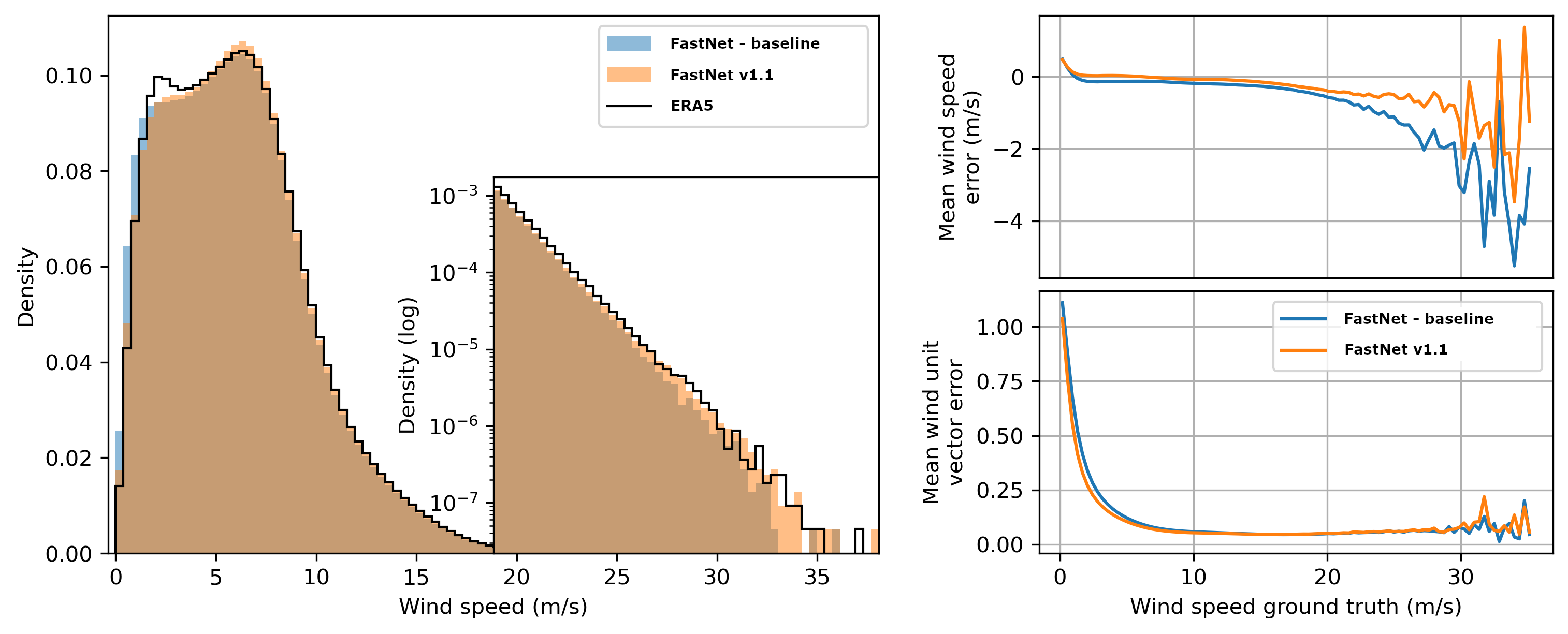}
\caption{As Figure \ref{fig:altwinds_dist} but for the FastNet v1.1 O96 model compared to baseline. 10 m wind speed at a 6-hour lead times evaluated over 2022.}
\label{fig:v1p1_wind_dist}
\end{figure*}

Of particular importance is that the benefits of the individual loss function modifications described in the preceding sections, are retained in the combined model. Figure \ref{fig:fastnet_v1p1_ciaran} shows that the reduction in mesh-scale artefacts in geostrophic winds from the inclusion of horizontal gradients remains effective when combined with MSH fine-tuning. Figure \ref{fig:v1p1_wind_dist} also demonstrates that the combined loss modifications produce significant improvements to the representation of wind speed and wind direction, and particularly for high wind speeds.

\section{Conclusions}
\label{sec:conclusions}

We present a set of experiments designed to assess the effects of alternative loss functions with the aim of improving the physical consistency of the output from MLWP models. The loss functions were developed and implemented in the FastNet MLWP model - a GNN based global prediction model that has a baseline performance that is competitive with leading MLWP and traditional NWP models on standard error metrics such as RMSE. The modifications were targetted at improving three specific deficiencies that are common to many deterministic MLWP models, namely, spectral bias that manifests as blurring of small-scale features, numerical artefacts that arise from the model architecture, and a slow bias to the forecast wind speeds.
We demonstrate that the modifications described here are effective in moving the model towards a more consistent physical representation of the atmosphere compared to the RMSE-only trained models. Moreover, the modifications are complementary and retain the desirable properties of individual changes when combined in a single model, which we present here as FastNet v1.1. Loss function design for MLWP is a simple and effective method of incorporating expert domain knowledge into MLWP models.

Some limitations of the present work should be noted:
\begin{itemize}
    \item The results shown here should be viewed as a proof-of-principle, and not as a optimal configurations for each of the loss functions. The loss function modifications introduce several new hyperparameters that require exploration, and it was beyond the scope of this work to find optimal settings, either individually or in combination. 
    \item The important role of the way the problem is formulated must be kept in mind. For example, several studies now show that the problem of spectral bias can be effectively removed by formulating the learning task probabilistically \citep{lang_aifs-crps_2024, price_probabilistic_2024}. We implement these ideas in a deterministic setting here for simplicity, but we would suggest that they are equally applicable to probabilistic models, where perhaps some of the trade-offs inherent in a deterministic setting would no longer apply.
\end{itemize}

In future development, we aim to address some of the limitations noted above. As MLWP models mature and move towards operational deployment, the importance of physical consistency of the output will be brought into sharp focus. Downstream products and services that rely on MLWP output will require spatio-temporal consistency and correct inter-variable relations to produce derived quantities, and as the output resolution of MLWP increases, the problem of spectral bias, or blurring, becomes even more pronounced. Efforts to tackle these issue now will pay dividends later by easing the transition of MLWP models from research to operations, building trust in their output, and extending their range of downstream uses.

\acknowledgments This work was funded by the Met Office and The Alan Turing Institute. We wish to thank the following people for their contributions to this work: Ryan Boult, Stephen Belcher, Maya Bronfeld, Mathew Corbett, Marc Deisenroth, Daniel Delbarre, Ben Fitzpatrick, Mark Girolami, Nicolas Guernion, Stephen Haddad, Richard Hattersley, Tom Henderson, Aaron Hopkinson, Richard Lawrence, Tomas Lazauskas, Clodagh Lynch, Ben MacArthur, Cyril Morcrette, Kelly O'Meara, Martin O'Reilly, Aled Owen, Joseph Palmer, James Penn, Bernat Puig-Camps, Christine Sheldon, Jonathan Starck, Hannah Sweeney, Richard Turner, Monica Vakil-Dewar, Luke Vinton, Simon Vosper, George Williams, Keith Williams.


\bibliographystyle{ametsocV6}
\bibliography{references}

\begin{thebibliography}{24}
\providecommand{\natexlab}[1]{#1}
\providecommand{\url}[1]{\texttt{#1}}
\renewcommand{\UrlFont}{\rmfamily}
\providecommand{\urlprefix}{URL }
\expandafter\ifx\csname urlstyle\endcsname\relax
  \providecommand{\doi}[1]{https://doi.org/\discretionary{}{}{}#1}\else
  \providecommand{\doi}{https://doi.org/\discretionary{}{}{}\begingroup \urlstyle{rm}\Url}\fi
\providecommand{\eprint}[2][]{\url{#2}}

\bibitem[{Baer(1972)}]{Baer1972}
Baer, F., 1972: An alternate scale representation of atmospheric energy spectra. \textit{Journal of Atmospheric Sciences}, \textbf{29~(4)}, 649 -- 664, \doi{10.1175/1520-0469(1972)029<0649:AASROA>2.0.CO;2}.

\bibitem[{Battaglia et~al.(2016)Battaglia, Pascanu, Lai, Jimenez~Rezende et~al.}]{battaglia2016interaction}
Battaglia, P., R.~Pascanu, M.~Lai, D.~Jimenez~Rezende, and Coauthors, 2016: Interaction networks for learning about objects, relations and physics. \textit{Advances in neural information processing systems}, \textbf{29}.

\bibitem[{Bi et~al.(2023)Bi, Xie, Zhang, Chen, Gu,, and Tian}]{bi_accurate_2023}
Bi, K., L.~Xie, H.~Zhang, X.~Chen, X.~Gu, and Q.~Tian, 2023: Accurate medium-range global weather forecasting with {3D} neural networks. \textit{Nature}, \textbf{619~(7970)}, 533--538, \doi{10.1038/s41586-023-06185-3}.

\bibitem[{Bonev et~al.(2023)Bonev, Kurth, Hundt, Pathak, Baust, Kashinath,, and Anandkumar}]{bonev_spherical_2023}
Bonev, B., T.~Kurth, C.~Hundt, J.~Pathak, M.~Baust, K.~Kashinath, and A.~Anandkumar, 2023: Spherical {Fourier} {Neural} {Operators}: {Learning} {Stable} {Dynamics} on the {Sphere}. arXiv, \urlprefix\url{http://arxiv.org/abs/2306.03838}, arXiv:2306.03838, \doi{10.48550/arXiv.2306.03838}.

\bibitem[{Bonev et~al.(2025)}]{bonev_fourcastnet_2025}
Bonev, B., and Coauthors, 2025: {FourCastNet} 3: {A} geometric approach to probabilistic machine-learning weather forecasting at scale. arXiv, \urlprefix\url{http://arxiv.org/abs/2507.12144}, arXiv:2507.12144 [cs], \doi{10.48550/arXiv.2507.12144}.

\bibitem[{Charlton-Perez et~al.(2024)}]{charlton-perez_ai_2024}
Charlton-Perez, A.~J., and Coauthors, 2024: Do {AI} models produce better weather forecasts than physics-based models? {A} quantitative evaluation case study of {Storm} {Ciarán}. \textit{npj Climate and Atmospheric Science}, \textbf{7~(1)}, 93, \doi{10.1038/s41612-024-00638-w}.

\bibitem[{Dahl et~al.(2014)Dahl, Dahl,, and Larsen}]{dahl2014surface}
Dahl, V.~A., A.~B. Dahl, and R.~Larsen, 2014: Surface detection using round cut. \textit{2014 2nd International Conference on 3D Vision}, IEEE, Vol.~2, 82--89.

\bibitem[{Daub et~al.(2025)}]{Daub_fastnet_paper1}
Daub, E.~G., and Coauthors, 2025: Technical overview and architecture of the fastnet machine learning weather prediction model, version 1.0.

\bibitem[{Ellis(2025)}]{Ellis2025}
Ellis, A.-L., 2025: Ml model architectures' power spectra characteristics and their relationships to ml model artefacts.

\bibitem[{Fornberg(2002)}]{Fornberg2002}
Fornberg, B., 2002: Calculation of weights in finite difference formulas. \textit{SIAM Review}, \textbf{40}, \doi{10.1142/9789814335867_0006}.

\bibitem[{Gage(1979)}]{Gage1979}
Gage, K.~S., 1979: Evidence for a k-5/3 law inertial range in mesoscale two-dimensional turbulence. \textit{Journal of Atmospheric Sciences}, \textbf{36~(10)}, 1950 -- 1954, \doi{10.1175/1520-0469(1979)036<1950:EFALIR>2.0.CO;2}.

\bibitem[{Hersbach et~al.(2020)}]{hersbach_era5_2020}
Hersbach, H., and Coauthors, 2020: The {ERA5} global reanalysis. \textit{Quarterly Journal of the Royal Meteorological Society}, \textbf{146~(730)}, 1999--2049, \doi{10.1002/qj.3803}.

\bibitem[{Keisler(2022)}]{keisler_forecasting_2022}
Keisler, R., 2022: Forecasting {Global} {Weather} with {Graph} {Neural} {Networks}. arXiv, \urlprefix\url{http://arxiv.org/abs/2202.07575}, arXiv:2202.07575 [physics].

\bibitem[{Kochkov et~al.()}]{kochkov_neural_2024}
Kochkov, D., and Coauthors, ????: Neural general circulation models for weather and climate. \textbf{632~(8027)}, 1060--1066, \doi{10.1038/s41586-024-07744-y}.

\bibitem[{Lam et~al.(2023)}]{lam_learning_2023}
Lam, R., and Coauthors, 2023: Learning skillful medium-range global weather forecasting. \textit{Science}, \textbf{382~(6677)}, 1416--1421, \doi{10.1126/science.adi2336}.

\bibitem[{Lang et~al.(2024{\natexlab{a}})}]{lang_aifs_2024}
Lang, S., and Coauthors, 2024{\natexlab{a}}: {AIFS} -- {ECMWF}'s data-driven forecasting system. arXiv, \urlprefix\url{http://arxiv.org/abs/2406.01465}, arXiv:2406.01465, \doi{10.48550/arXiv.2406.01465}.

\bibitem[{Lang et~al.(2024{\natexlab{b}})}]{lang_aifs-crps_2024}
Lang, S., and Coauthors, 2024{\natexlab{b}}: {AIFS}-{CRPS}: {Ensemble} forecasting using a model trained with a loss function based on the {Continuous} {Ranked} {Probability} {Score}. arXiv, \urlprefix\url{http://arxiv.org/abs/2412.15832}, arXiv:2412.15832, \doi{10.48550/arXiv.2412.15832}.

\bibitem[{Pathak et~al.(2022)}]{pathak_fourcastnet_2022}
Pathak, J., and Coauthors, 2022: {FourCastNet}: {A} {Global} {Data}-driven {High}-resolution {Weather} {Model} using {Adaptive} {Fourier} {Neural} {Operators}. \urlprefix\url{https://arxiv.org/abs/2202.11214v1}.

\bibitem[{Pfaff et~al.(2021)Pfaff, Fortunato, Sanchez-Gonzalez,, and Battaglia}]{pfaff2021learning}
Pfaff, T., M.~Fortunato, A.~Sanchez-Gonzalez, and P.~W. Battaglia, 2021: Learning mesh-based simulation with graph networks. \textit{International Conference on Learning Representations (ICLR 2021)}.

\bibitem[{Price et~al.(2024)}]{price_probabilistic_2024}
Price, I., and Coauthors, 2024: Probabilistic weather forecasting with machine learning. \textit{Nature}, 1--7, \doi{10.1038/s41586-024-08252-9}.

\bibitem[{Rasp et~al.(2020)Rasp, Dueben, Scher, Weyn, Mouatadid,, and Thuerey}]{rasp_weatherbench_2020}
Rasp, S., P.~D. Dueben, S.~Scher, J.~A. Weyn, S.~Mouatadid, and N.~Thuerey, 2020: {WeatherBench}: {A} {Benchmark} {Data} {Set} for {Data}‐{Driven} {Weather} {Forecasting}. \textit{Journal of Advances in Modeling Earth Systems}, \textbf{12~(11)}, e2020MS002\,203, \doi{10.1029/2020MS002203}.

\bibitem[{Rasp and Thuerey(2021)Rasp, and Thuerey}]{rasp_datadriven_2021}
Rasp, S., and N.~Thuerey, 2021: Data‐{Driven} {Medium}‐{Range} {Weather} {Prediction} {With} a {Resnet} {Pretrained} on {Climate} {Simulations}: {A} {New} {Model} for {WeatherBench}. \textit{Journal of Advances in Modeling Earth Systems}, \textbf{13~(2)}, e2020MS002\,405, \doi{10.1029/2020MS002405}.

\bibitem[{Subich et~al.(2025)Subich, Husain, Separovic,, and Yang}]{subich_fixing_2025}
Subich, C., S.~Z. Husain, L.~Separovic, and J.~Yang, 2025: Fixing the {Double} {Penalty} in {Data}-{Driven} {Weather} {Forecasting} {Through} a {Modified} {Spherical} {Harmonic} {Loss} {Function}. arXiv, \urlprefix\url{http://arxiv.org/abs/2501.19374}, arXiv:2501.19374, \doi{10.48550/arXiv.2501.19374}.

\bibitem[{Wilson(2025)}]{wilson_deep_2025}
Wilson, A.~G., 2025: Deep {Learning} is {Not} {So} {Mysterious} or {Different}. arXiv, \urlprefix\url{http://arxiv.org/abs/2503.02113}, arXiv:2503.02113 [cs], \doi{10.48550/arXiv.2503.02113}.

\end{thebibliography}

\includepdf{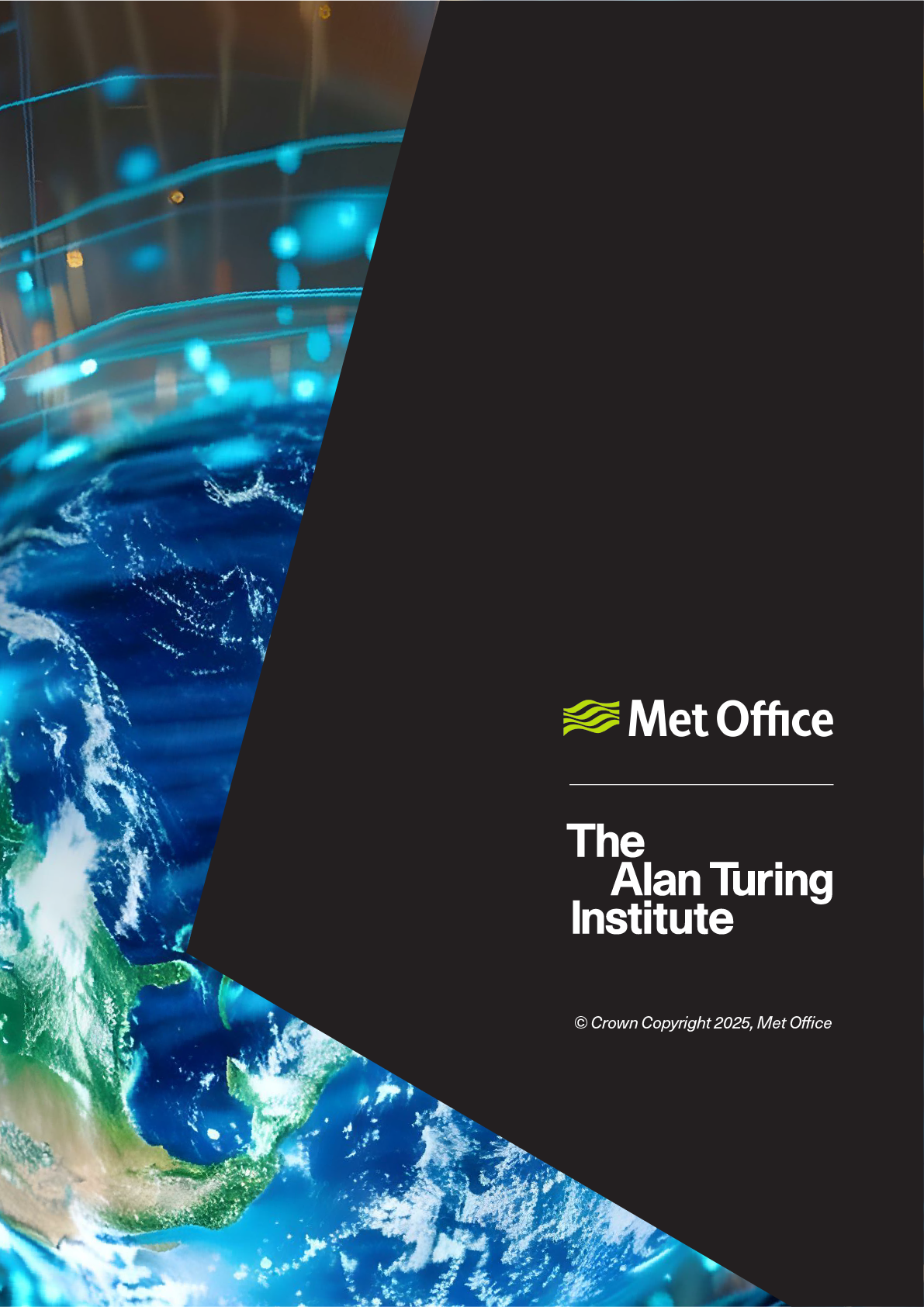}

\end{document}